\documentclass[superscriptaddress, nofootinbib, longbibliography, amsmath, amssymb, twocolumn]{revtex4-1}
\usepackage[margin=1in]{geometry} 
\usepackage[english]{babel}
\usepackage[utf8]{inputenc}
\newcounter{TaskCounter}
\usepackage{graphicx}
\usepackage{xspace}
\usepackage{siunitx}
\DeclareSIUnit\angstrom{\text {Å}}
\usepackage{orcidlink}
\usepackage{hyperref}
\DeclareRobustCommand{\gobblefive}[5]{}
\newcommand*{\SkipTocEntry}{\addtocontents{toc}{\gobblefive}}

\newcommand*{\smiles}{{S}{\footnotesize MILES}\xspace}
\newcommand*{\deepsmiles}{{D}{\footnotesize EEP}{S}{\footnotesize MILES}\xspace}
\newcommand*{\selfies}{{S}{\footnotesize ELFIES}\xspace}
\newcommand*{\inchi}{{I}{\footnotesize N}{C}{\footnotesize H}{I}\xspace}

\newcommand*{\helm}{\textsc{Helm}\xspace}

\hypersetup{hidelinks,colorlinks=true,breaklinks=true,urlcolor=blue,allcolors=blue,pdftitle={Title},pdfauthor={Author}}

\usepackage{listings, xcolor}
\definecolor{codegreen}{rgb}{0,0.6,0}
\definecolor{codegray}{rgb}{0.5,0.5,0.5}
\definecolor{codepurple}{rgb}{0.58,0,0.82}
\definecolor{tqblue}{HTML}{08293d}
\definecolor{backcolour}{HTML}{fefdf5}

\lstdefinestyle{pythonstyle}{
    backgroundcolor=\color{backcolour},   
    commentstyle=\color{codegreen},
    keywordstyle=\color{magenta},
    numberstyle=\tiny\color{codegray},
    stringstyle=\color{codepurple},
    basicstyle=\ttfamily\footnotesize\color{tqblue},
    breakatwhitespace=false,         
    breaklines=true,
    postbreak=\mbox{\textcolor{magenta}{$\hookrightarrow$}\space},                 
    captionpos=b,                    
    keepspaces=true,                 
    numbers=left,                    
    numbersep=5pt,                  
    showspaces=false,                
    showstringspaces=false,
    showtabs=false,                  
    tabsize=2
}

\UseRawInputEncoding 
\begin{document}

\title{SELFIES and the future of molecular string representations}

\author{Mario~Krenn~\orcidlink{0000-0003-1620-9207 }}
\email{mario.krenn@mpl.mpg.de}
\affiliation{Max Planck Institute for the Science of Light (MPL), Erlangen, Germany.}

\author{Qianxiang~Ai~\orcidlink{0000-0002-5487-2539}}
\affiliation{Department of Chemistry, Fordham University, The Bronx, New York, USA.}

\author{Senja~Barthel~\orcidlink{0000-0002-9175-5067}}
\affiliation{Department of Mathematics, Vrije Universiteit Amsterdam, Amsterdam, the Netherlands.}

\author{Nessa~Carson~\orcidlink{0000-0002-2769-1775}}
\affiliation{Syngenta Jealott's Hill International Research Centre, Bracknell, Berkshire, UK.}

\author{Angelo~Frei~\orcidlink{0000-0001-6169-2491}}
\affiliation{Department of Chemistry, Imperial College London, Molecular Sciences Research Hub, White City Campus, Wood Lane, London, UK.}

\author{Nathan~C.~Frey~\orcidlink{0000-0001-5291-6131}}
\affiliation{Massachusetts Institute of Technology, Cambridge, Massachusetts, USA.}

\author{Pascal~Friederich~\orcidlink{0000-0003-4465-1465}}
\affiliation{Institute of Theoretical Informatics, Karlsruhe Institute of Technology, Karlsruhe, Germany.}
\affiliation{Institute of Nanotechnology, Karlsruhe Institute of Technology, Eggenstein-Leopoldshafen, Germany.}

\author{Th\'{e}ophile~Gaudin~\orcidlink{0000-0003-3914-854X}}
\affiliation{Department of Computer Science, University of Toronto, Canada.}
\affiliation{IBM Research Europe, Z\"urich, Switzerland.}

\author{Alberto~Alexander~Gayle~\orcidlink{0000-0002-9805-8440}}
\affiliation{Unaffiliated; Sapporo, Japan.}

\author{Kevin~Maik~Jablonka~\orcidlink{0000-0003-4894-4660}}
\affiliation{Laboratory of Molecular Simulation (LSMO), Institut des Sciences et Ing\'{e}nierie Chimiques, Ecole Polytechnique F\'{e}d\'{e}rale de Lausanne (EPFL), Sion, Valais, Switzerland.}

\author{Rafael~F.~Lameiro~\orcidlink{0000-0003-4466-2682}}
\affiliation{Medicinal and Biological Chemistry Group, S\~{a}o Carlos Institute of Chemistry, University of S\~{a}o Paulo, S\~{a}o Paulo, Brazil.}

\author{Dominik~Lemm~\orcidlink{0000-0002-8075-1765}}
\affiliation{Faculty of Physics, University of Vienna, Vienna, Austria.}

\author{Alston~Lo~\orcidlink{0000-0003-1744-1446}}
\affiliation{Department of Computer Science, University of Toronto, Canada.}

\author{Seyed~Mohamad~Moosavi~\orcidlink{0000-0002-0357-5729}}
\affiliation{Department of Mathematics and Computer Science, Freie Universit\"{a}t Berlin, Berlin, Germany.}

\author{Jos\'e~Manuel~N\'apoles-Duarte~\orcidlink{0000-0001-6823-4733}}
\affiliation{Facultad de Ciencias Qu\'imicas, Universidad Aut\'onoma de Chihuahua, Chihuahua, Mexico.}

\author{AkshatKumar~Nigam~\orcidlink{0000-0002-5152-2082}}
\affiliation{Department of Computer Science, Stanford University, California, USA.}

\author{Robert~Pollice~\orcidlink{0000-0001-8836-6266}}
\affiliation{Department of Computer Science, University of Toronto, Canada.}
\affiliation{Chemical Physics Theory Group, Department of Chemistry, University of Toronto, Canada.}

\author{Kohulan~Rajan~\orcidlink{0000-0003-1066-7792}}
\affiliation{Institute for Inorganic and Analytical Chemistry, Friedrich-Schiller Universit\"{a}t Jena, Jena, Germany.}

\author{Ulrich~Schatzschneider~\orcidlink{0000-0002-1960-1880}}
\affiliation{Institut f\"ur Anorganische Chemie, Julius-Maximilians-Universit\"{a}t W\"urzburg, W\"urzburg, Germany.}

\author{Philippe~Schwaller~\orcidlink{0000-0003-3046-6576}}
\affiliation{Laboratory of Artificial Chemical Intelligence (LIAC), Institut des Sciences et Ing\'{e}nierie Chimiques, Ecole Polytechnique F\'{e}d\'{e}rale de Lausanne (EPFL), Lausanne, Switzerland.}
\affiliation{National Centre of Competence in Research (NCCR) Catalysis, Ecole Polytechnique F\'{e}d\'{e}rale de Lausanne (EPFL), Lausanne, Switzerland.}
\affiliation{IBM Research Europe, Z\"urich, Switzerland.}

\author{Marta~Skreta~\orcidlink{0000-0001-6831-8398}} 
\affiliation{Department of Computer Science, University of Toronto, Canada.}
\affiliation{Vector Institute for Artificial Intelligence, Toronto, Canada.}

\author{Berend~Smit~\orcidlink{0000-0003-4653-8562}}
\affiliation{Laboratory of Molecular Simulation (LSMO), Institut des Sciences et Ing\'{e}nierie Chimiques, Ecole Polytechnique F\'{e}d\'{e}rale de Lausanne (EPFL), Sion, Valais, Switzerland.}

\author{Felix~Strieth-Kalthoff~\orcidlink{0000-0003-1357-5500}}
\affiliation{Chemical Physics Theory Group, Department of Chemistry, University of Toronto, Canada.}

\author{Chong~Sun~\orcidlink{0000-0002-8299-9094}}
\affiliation{Department of Computer Science, University of Toronto, Canada.}

\author{Gary~Tom~\orcidlink{0000-0002-8470-6515}}
\affiliation{Chemical Physics Theory Group, Department of Chemistry, University of Toronto, Canada.}

\author{Guido~Falk~von~Rudorff~\orcidlink{0000-0001-7987-4330}}
\affiliation{Faculty of Physics, University of Vienna, Vienna, Austria.}

\author{Andrew~Wang~\orcidlink{0000-0003-3647-500}}
\affiliation{Chemical Physics Theory Group, Department of Chemistry, University of Toronto, Canada.}
\affiliation {Solar Fuels Group, Department of Chemistry, University of Toronto, Canada}

\author{Andrew~White~\orcidlink{0000-0002-6647-3965}}
\affiliation{Department of Chemical Engineering, University of Rochester, USA.}

\author{Adamo~Young~\orcidlink{0000-0001-5620-6241}}
\affiliation{Department of Computer Science, University of Toronto, Canada.}
\affiliation{Vector Institute for Artificial Intelligence, Toronto, Canada.}

\author{Rose~Yu~\orcidlink{0000-0002-8491-7937}}
\affiliation{Department of Computer Science and Engineering, University of California, San Diego, USA.}

\author{Al\'an~Aspuru-Guzik\orcidlink{0000-0002-8277-4434}}
\email{alan@aspuru.com}
\affiliation{Department of Computer Science, University of Toronto, Canada.}
\affiliation{Chemical Physics Theory Group, Department of Chemistry, University of Toronto, Canada.}
\affiliation{Vector Institute for Artificial Intelligence, Toronto, Canada.}
\affiliation{Department of Chemical Engineering and Applied Chemistry, University of Toronto, Canada.}
\affiliation{Department of Materials Science, University of Toronto, Canada.}
\affiliation{Canadian Institute for Advanced Research (CIFAR) Lebovic Fellow, Toronto, Canada.}

\begin{abstract}
Artificial intelligence (AI) and machine learning (ML) are expanding in popularity for broad applications to challenging tasks in chemistry and materials science. Examples include the prediction of properties, the discovery of new reaction pathways, or the design of new molecules. The machine needs to read and write fluently in a chemical language for each of these tasks. Strings are a common tool to represent molecular graphs, and the most popular molecular string representation, \smiles, has powered cheminformatics since the late 1980s. However, in the context of AI and ML in chemistry, \smiles has several shortcomings -- most pertinently, most combinations of symbols lead to invalid results with no valid chemical interpretation. To overcome this issue, a new language for molecules was introduced in 2020 that guarantees 100\% robustness: \selfies (SELF-referencIng Embedded Strings). \selfies has since simplified and enabled numerous new applications in chemistry. In this manuscript, we look to the future and discuss molecular string representations, along with their respective opportunities and challenges. We propose 16 concrete Future Projects for robust molecular representations. These involve the extension toward new chemical domains, exciting questions at the interface of AI and robust \textit{languages}, and interpretability for both humans and machines. We hope that these proposals will inspire several follow-up works exploiting the full potential of molecular string representations for the future of AI in chemistry and materials science.
\end{abstract}

\maketitle

\newpage
\tableofcontents
\section{Introduction}
The discovery of new materials and molecules with exceptional properties could lead to enormous scientific, technological and ultimately societal impact. In the last few years, digital discoveries -- that is, \textit{in\,silico} discoveries using computers -- have been significantly reinforced through ML applications and other AI tools for chemistry. Specifically, recent advances in AI and ML have sparked numerous new applications in quantum chemistry \cite{zubatiuk2021development,huang2021ab,behler2021four,westermayr2021machine,keith2021combining,dral2021molecular,von2020exploring}, molecular dynamics simulations \cite{glielmo2021unsupervised,unke2021machine,friederich2021machine}, prediction of molecular properties \cite{walters2021applications,deringer2021gaussian,nandy2021computational} and reactivity \cite{gallegos2021importance,zuranski2021predicting,meuwly2021machine,jorner2021organic}, artificial molecular design \cite{sanchez2018inverse,terayama2021black,janet2021navigating,pollice2021data,whiteDeep2021}, and the formulation of design heuristics \cite{crawford2021data,jablonka2020big} and representations \cite{warr2011representation, wigh2022review}. One germane question in all these applications is: which language should be used to symbolically represent molecules and materials?

Since the 1980s, \smiles (Simplified Molecular Input Line Entry System) strings have been a very prominent graph representation in computational chemistry. However, questions have arisen as to whether \smiles is an ideal language for computer applications that are tasked to discover new structures. For example, \smiles are not robust on their own, which means that generative models are likely to create strings that do not represent valid molecular graphs. A large body of work has been devoted to resolving this issue in recent years. Much of the advances came from model-dependent solutions, fixing the problem inside ML algorithms \cite{jin2018junction,popova2018deep}.

In 2020, some of us introduced \selfies\cite{krenn2020self}\footnote{\selfies can be installed via \texttt{pip install selfies}\\
\url{https://github.com/aspuru-guzik-group/selfies}}. This new string-based representation circumvents the issue of robustness by defining a formal grammar that always leads to a valid molecular graph. This new molecular graph representation has simplified numerous applications in cheminformatics and even enabled new ones. Given this exciting potential, the authors assembled\footnote{The authors assembled in a virtual mini-workshop in August 2021 organized by IOP and the Acceleration Consortium, on the topic of this paper.} to jointly discuss the future of \selfies in terms of generalizations and new applications. Here, we present an overview of the progress as well as outstanding questions, formulating 16 concrete projects and challenging ideas for the next years.

The manuscript is structured as follows: we first summarize briefly the 250-year-long history of molecular representations. Then we look at modern representations and discuss their strengths and weaknesses. This motivates a look into the future, where many open questions remain. In our journey, we also visit stochastic macromolecules and crystals. We will go further down the rabbit hole of inorganic chemistry and look at the potential for modeling and predicting chemical reactions. Then, we analyse the performance of string-based and non-string-based representations in terms of ML, and finally also investigate questions about general interpretability of chemical languages -- for both human and artificial scientists. During our journey through different fields of chemistry and AI research, we propose 16 independent stand-alone research projects which could define the future of molecular representations for AI in chemistry.

\section{Historical review}
Shaping the future of molecular representation is only sensible if we comprehend its history. Here, we briefly describe the 250-year evolution of chemical notations and the advent of modern string representations for molecules. Detailed accounts of the history can be found in other papers \cite{wsl1952,Dorward1965,Fletcher1974,warr1982,Hepler2015,Fauque2019}.

\textbf{1787:} The origin of chemical nomenclature is rooted in the seminal work ``{\em M\'ethode de nomenclature chimique}'', with contributions from Lavoisier and others \cite{Lavoisier1787}. This work ushered in the modern, post-alchemy era of chemical nomenclature.

\textbf{1808:} Dalton developed his atomic theory and used symbols to represent elements and compounds \cite{Dalton1808}. These symbols resembled those used in the prior, alchemical era. For example, the elements hydrogen and sulfur were represented by $\bigodot$ and $\bigoplus$, respectively, while the compound water was represented as $\bigodot \bigcirc$. However, such highly specialized symbols had two major drawbacks. Firstly, they were non-intuitive and therefore cumbersome for others to learn and apply. Secondly, they were incompatible with contemporaneous printing methods, resulting in limited circulation of Dalton's work. 

\textbf{1813:} Berzelius sought to address this by proposing a terminology where the first letters of the Latin names of a substance were used in lieu of symbols \cite{Berzelius1813}. This new notation represented chemical ratios rather than molecular structures. 

\textbf{1889--1911:} International committees were formed to standardize the chemical nomenclature. The International Chemistry Committee published the \textit{Geneva Rules for Organic Chemistry} in 1889. This was the first attempt to standardize chemical nomenclature \cite{Hepler2015}. Nomenclature reforms continued with the International Association of Chemical Societies, which convened in 1911 in Paris. However, the proceedings were interrupted by the outbreak of World War I \cite{IACS1912}. 

\textbf{1919--1930:} The International Union of Pure and Applied Chemistry (IUPAC) was formed following the conclusion of World War I. In 1921, the Union continued to advance chemical nomenclature, culminating in 1930 with the so-called {\em Li\`{e}ge Rules} \cite{Fauque2019}. 

\textbf{1944--1947:} While the outbreak of World War II interrupted the work of IUPAC, Dyson independently published a seminal work entitled ``A Notation for Organic Compounds'' in 1944 \cite{Dyson1944}. A revised version: ``A New Notation and Enumeration System for Organic Compounds.'' was subsequently accepted by IUPAC in 1947 \cite{Fletcher1974,Dyson1947}. The latter received criticism for not adding to the problem of chemical nomenclature, and that better explanations would be found in the original lecture in 1944. The claims in Dyson's work were taken with reservations, especially the affirmation that there was only one possible cipher for any one chemical compound when there was not enough evidence and little scrutiny by the chemistry community \cite{Brightman1947ANN}. There was a feeling that he was prescribing a sledge-hammer to crush a nut. 

\begin{figure}[htbp!]
\centering
\includegraphics[width=0.48\textwidth]{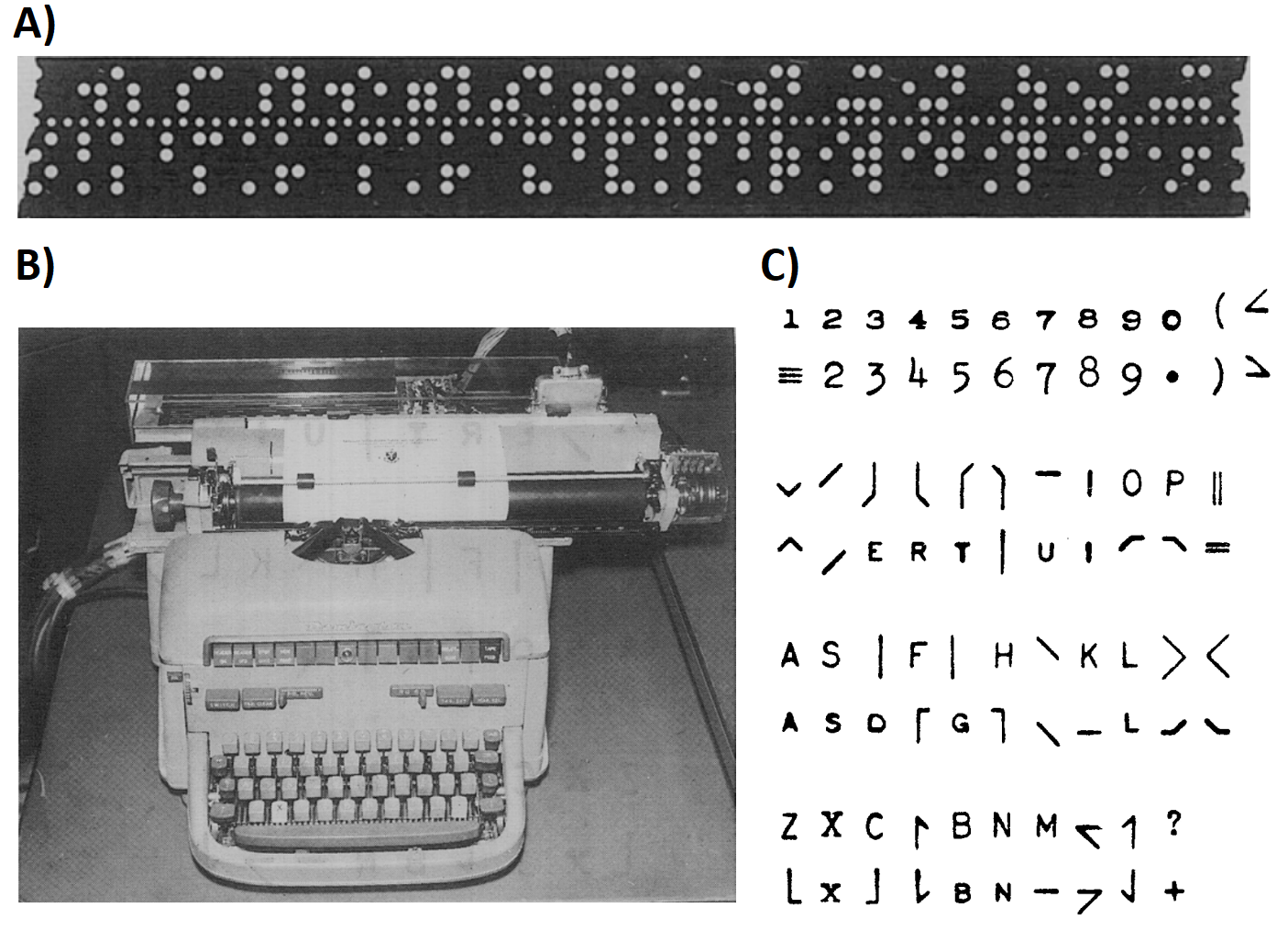}
\caption{ (A) Typical tape obtained with the Army Chemical Typewriter (ACT) built by members of the Walter Reed Army Institute of Research. (B) The ACT, a mechanical typewriter for the encoding of chemical structures. (C) Typed characters from the ACT. Image from Reference \cite{feldman63}.}
\label{fig:typew}
\end{figure} 

\textbf{1949--1951:} With the advent of computers, there was a new necessity to adapt chemical formulas to line notation using ASCII, thereby eliminating, among other features, the use of subscript and Greek letters \cite{raos2012}. In 1949, the IUPAC Commission on Codification, Ciphering, and Punched Card Techniques opened a call for proposals regarding an international notation system. The criteria for the proposed annotation system included simplicity of use and ease of printing and typewriting. In 1951, the commission reviewed line notations with contributions from seven different proposals \cite{cen1952}. From those, Dyson's ciphering remained the standard, though many alternatives were used in practice. Among these, the Wiswesser Line Notation (WLN) \cite{wsl1952} is the most noteworthy. It provided a ``compact way of uniquely and unambiguously representing the complete topology of a chemical molecule'' and was preferred by scientists for many decades thereafter \cite{warr1982}.


\textbf{1961--1969:}
During this era, the WLN method became the \textit{de facto} standard in computer and punched card approaches to storing large data sets of chemical compounds \cite{wis1982}. Subsequent efforts focused on automated hardware specially designed to codify molecules, like the Army Chemical Typewriter (Figure~\ref{fig:typew}), or, alternatively, on improving machine readability and storage capacity, for example the Hayward Notation (1961) \cite{Hayward1961} and the Skolnik Notation (1969) \cite{skol1964}. In the former, the aim was to establish a basis for a one-to-one relationship between structure, cipher, and nomenclature, while for the latter it was to have the notations conform to the accepted chemical structures, and to invoke relatively few rules.

\section{Modern molecular string representations}
\begin{figure*}[htbp!]
\centering
\includegraphics[width=0.85\textwidth]{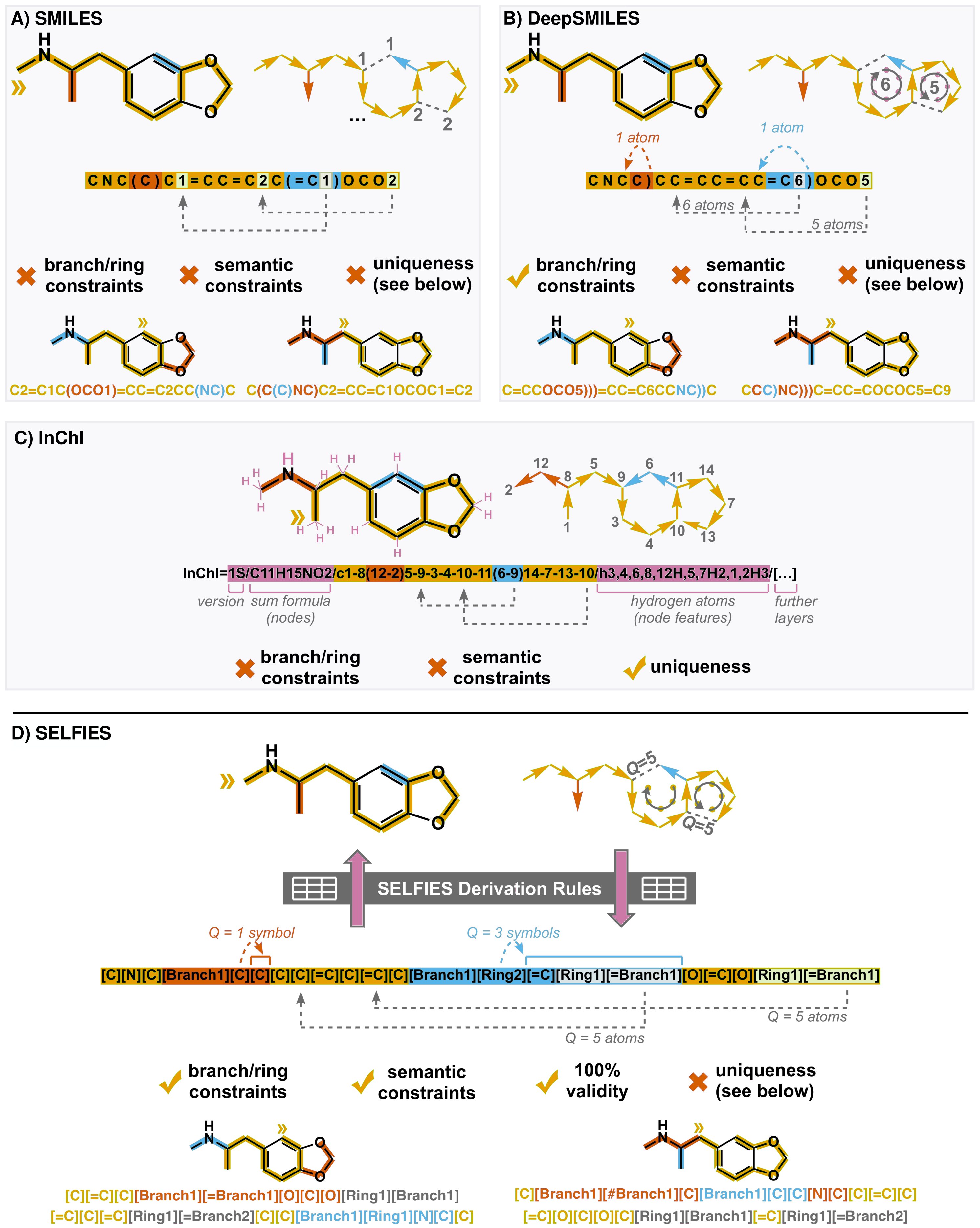}
\caption{Derivation of established string representations (A) \smiles, (B) \deepsmiles, and (C) \inchi from molecular structures, using 3,4-methylenedioxymethamphetamine (MDMA) as example. Branches and ring closures are represented by specific syntax based on the main path (orange). (D) Derivation of a \selfies string from the molecular structure, building on the corresponding derivation rules.} 
\label{fig:stringreps_general}
\end{figure*}

The development of molecular string representations has continued in the direction laid out by \mbox{IUPAC} in 1949. However, advances in computer power and cheminformatics applications have accelerated development far beyond the use cases originally envisioned. In the following section, we discuss four molecular string representations which are widely used today, with a focus on their applications in AI for chemistry and material science.

\subsection{SMILES}
Weininger published \smiles in 1988 with the goal to serve the needs of ``modern chemical information processing'' \cite{weininger1988smiles,weininger1989smiles}. The development of \smiles focused on the implementation of molecular graph theory, to allow for rigorous structure specification with a grammar that is both minimal and natural. \smiles has since become the \textit{de facto} standard representation in cheminformatics.

An example for the \smiles representation is shown in Figure~\ref{fig:stringreps_general}. In \smiles, molecules are defined as a chain of atoms which are written as letters in a string. Branches in the molecule are defined within parentheses, while ring closures are indicated by two matching numbers. The \smiles grammar, though simple, allows for the description of complex structures as well as properties such as stereochemistry, aromatic bonds, chirality, ions, and isotopes.

While \smiles has been a workhorse for cheminformatics over the last three decades, in recent years, new applications in cheminformatics have exposed several weaknesses which motivated the introduction of new molecular string representations. Firstly, multiple different \smiles strings can represent the same molecule (e.g., see Figure~\ref{fig:stringreps_general}a). This weakness has been addressed by a different representation called \inchi, which we will explain below, and can be enforced by post-processing \textit{canonicalization} via tools such as RDKit \cite{landrum2013rdkit}.

Another weakness is that \smiles has no mechanism to ensure that molecular strings are valid with respect to syntax and physical principles. An example of the former is \texttt{CC(CCCC}, a string with an unpaired open parenthesis. This string has no valid interpretation as a molecular graph. Semantic errors involve strings that form valid graphs, but do not reflect valid chemical structures. For example, the string \texttt{CO=CC} represents a molecular graph with an oxygen atom that has three bonds -- a violation of the maximum number of bonds that neutral oxygen can form. 

The lack of syntactic and semantic robustness has a significant impact with respect to the validity of computer-designed molecules based on evolutionary or deep learning methods \cite{schneiderComputerbasedNovoDesign2005a,gomez2018automatic,sanchez2018inverse}. One solution has been the design of special ML models that attempt to enforce robustness \cite{jin2018junction,ma2018constrained,liu2018constrained}. A more fundamental solution is the modification of the molecular representation itself. O'Boyle and Dalke pioneered this approach by developing \deepsmiles, a modification of \smiles which obviates most syntactic errors, while semantic mistakes were still possible \cite{o2018deepsmiles}. Finally, 2020 witnessed the release of \selfies -- a molecular string representation \cite{krenn2020self} that is 100\% robust to both syntactic and semantic errors.

\subsection{\inchi}
\smiles are not unique representations of molecular graphs, i.e., a structure can be represented by multiple strings and custom identifiers. This makes it difficult to construct large-scale databases where each structure has to map to a unique label and \textit{vice versa}. The International Chemical Identifier (\inchi) was created in 2013 by IUPAC as an open-source software to encode molecular structures in order to standardize searching across databases and the internet \cite{heller2013inchi}. \inchi strings are composed of six main layers and multiple sublayers, where each layer represents a specific category of information about the molecule (sublayers include chemical formula, atomic connections, charges, stereochemistry). There are a number of advantages introduced by the \inchi syntax. The first is that molecules have a canonical representation, which allows straightforward linking in databases. O'Boyle created a method based on this feature of \inchi that generates universal \smiles strings to standardize the output from different cheminformatics toolkits \cite{o2012towards}. Another benefit of \inchi is that the layered structure encodes hierarchical information, and so two molecules which are derivatives of each other will have the same parent structure. Finally, \inchi is more expressive than \smiles and is able to encode more information. For example, \inchi can specify which hydrogen atoms are mobile and which are immobile \cite{heller2013inchi}. This allows for tautomers of the same molecule to be represented by the same \inchi string, while with the \smiles framework, each tautomer is represented by a different string. Also, \smiles requires explicit notation of double bond locations, while \inchi infers them. Consequently, resonance structures are represented by a single \inchi string but potentially multiple \smiles strings. There are also a number of disadvantages with the use of \inchi strings. The first is that the hierarchical structure and syntax makes the notation difficult to read by humans (although this is a point of contention as the readability improves with usage, we come back to this aspect in Section IX). The complicated syntax also makes it more difficult to employ \inchi in generative modeling, as there are a number of arithmetic and grammatical rules that are difficult to enforce when sampling a new molecule from deep learning models. Moreover, the current standard \inchi consistently disconnects bonds to metal atoms, which leads to loss of important stereochemical and bonding information. However, this behavior might change in future versions \cite{goodman2021inchi}. In practice, it has been found that \inchi performs worse than \smiles in ML-based applications, likely due to the above-mentioned reasons \cite{gomez2018automatic}.

\subsection{\deepsmiles}
Deep neural networks are increasingly used to create generative models for the design of new molecules \cite{sanchez2018inverse}. Many models were trained using molecules encoded as \smiles strings. These models are subsequently queried to generate \smiles strings representing molecules with specific target properties. However, the resulting \smiles may have unmatched parentheses or ring closure symbols, rendering the molecule invalid. To resolve these issues, O'Boyle and Dalke created \deepsmiles, which encodes \smiles into a syntax more suitable for automated inverse design such as deep generative models \cite{o2018deepsmiles}. The \deepsmiles grammar only uses one symbol to represent ring closures (instead of two). This symbol is a number which indicates how far back in the string the ring is connected. Branching is represented by one or more closing parentheses, where the number indicates branch length. Thereby, \deepsmiles resolves most cases of syntactical mistakes. This advance leads to greater robustness compared to \smiles with respect to random mutations and deep generative models \cite{krenn2020self}. However, \deepsmiles strings still allow for semantically incorrect strings, i.e.,\ molecules that violate basic physical constraints. This factor points to a need for an even more robust molecular grammar.

\subsection{\selfies}
Introduced in 2020, \selfies is a 100\% robust molecular string representation \cite{krenn2020self}. That is, \selfies cannot produce an invalid molecule, as every combination of symbols in the \selfies alphabet maps to a chemically valid graph. Let us imagine the same for a natural language, such as English. In the overwhelming majority of cases, an arbitrary combination of letters from the Latin alphabet (a--z) will not lead to a valid word. In this sense, English is not robust, while \selfies is robust with respect to chemistry.
\begin{figure}[htbp!]
\centering
\includegraphics[width=0.48\textwidth]{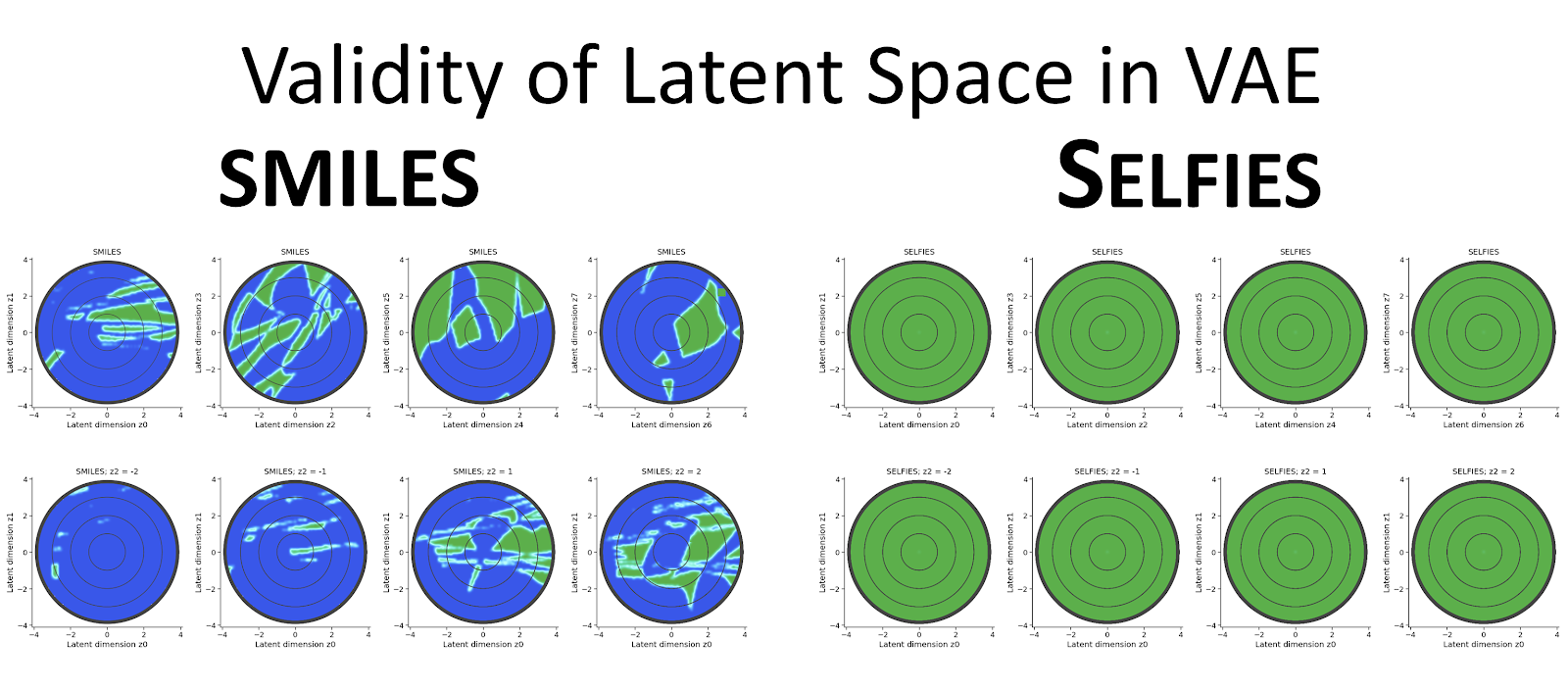}
\caption{Decoding points from the internal representation (latent space) of a Variational AutoEncoder (VAE). Green stands for valid and blue for invalid molecules. The left image is trained using \smiles strings, most of its latent space representing invalid molecular strings. The right image shows the latent space of a VAE trained with \selfies. Every point stands for a physically meaningful molecule. Figure from \cite{krenn2020self}.}
\label{fig:selfies_latentspace}
\end{figure}

\selfies is a formal grammar (or automaton) with derivation rules. This can be understood as a small computer program with minimal memory to achieve 100\% robust derivation. The \selfies grammar is designed with the explicit aim of eliminating syntactically and semantically invalid molecules, for example in generative tasks.

In \smiles, syntactic invalidity consists of unbalanced parentheses or ring identifiers. For instance, a generative model using \smiles may generate a string that includes an open parenthesis with no corresponding closing parenthesis. The resulting string would represent an invalid graph. The problem stems from the non-local definition of rings and branches, which has already been addressed through the introduction of \deepsmiles \cite{o2018deepsmiles}. To resolve these issues, \selfies follows a different approach. Here, rings and branches are both defined at one single location. Special symbols (such as \texttt{[Branch1]} or \texttt{[Ring1]}) start a branch or ring. Instead of using an end-symbol, the subsequent token in the string defines the length of the branch or ring. To achieve that, the next symbol is \textit{overloaded} (similar to function overloading in programming languages allowing creation of multiple functions with identical names but different implementations) by a number (see the concrete overloading list of \selfies version 2.0 in Table \ref{selfiesOverloading}). With these ideas, all syntactic mistakes are resolved.

\begin{table}[htbp!]
	\centering
	\caption{ List of \selfies symbols which are overloaded with numeric values if they appear after a ring or branch token. It is a hexadecimal system and larger numbers can be represented by overloading the next n symbols.}
	\renewcommand*{\arraystretch}{1.3} 
	\begin{tabular}{|cl|cl|}
	\hline
	Index & Symbol               & Index & Symbol               \\ \hline
	0     & \texttt{[C]}         & 8     & \texttt{[\#Branch2]} \\
	1     & \texttt{[Ring1]}     & 9     & \texttt{[O]}         \\
	2     & \texttt{[Ring2]}     & 10    & \texttt{[N]}         \\
	3     & \texttt{[Branch1]}   & 11    & \texttt{[=N]}        \\
	4     & \texttt{[=Branch1]}  & 12    & \texttt{[=C]}        \\
	5     & \texttt{[\#Branch1]} & 13    & \texttt{[\#C]}       \\
	6     & \texttt{[Branch2]}   & 14    & \texttt{[S]}         \\
	7     & \texttt{[=Branch2]}  & 15    & \texttt{[P]}         \\ \hline
	\multicolumn{4}{|l|}{All other symbols are assigned index 0.}  \\ \hline
	\end{tabular}
	\label{selfiesOverloading}
\end{table}

Semantic mistakes lead to molecular graphs that violate physical constraints. They are avoided by applying another concept from theoretical computer science -- formal grammar or formal automata \cite{hopcroft2001introduction}. The formal automaton derives the molecules, and every derivation step can change the state of the automaton. As the state defines the rules for the next derivation step, it can be used as a minimal memory that encodes physical constraints and ensures that only meaningful molecules are derived. \selfies can be seen as a very simple programming language for chemistry, and a \selfies string is a program that creates a valid molecular graph upon execution. This leads to interesting consequences and possibilities, which we will discuss in Section VIII.

Robustness can be demonstrated by inspecting the internal latent space of a deep learning model that is trained once with \smiles and once with \selfies (Figure~\ref{fig:selfies_latentspace}). Without changing anything inside the ML model, every \selfies output is physically valid. Not surprisingly, \selfies has already been shown to improve, simplify, or even enable new AI-driven applications in cheminformatics. These include genetic algorithms \cite{nigam2019augmenting}, curiosity-based exploration \cite{thiede2020curiosity}, efficient combinatorial methods \cite{nigam2021beyond}, and many other topics to be discussed later.

The library contains two core functions that facilitate the translation between \smiles and \selfies representations, alongside other peripheral functions for manipulating \selfies strings. The following depicts a simple use case of \selfies: 
\begin{lstlisting}[language=Python]
import selfies as sf

benzene = "c1ccccc1" 

# SMILES to SELFIES
benzene_sf = sf.encoder(benzene)  
# [C][=C][C][=C][C][=C][Ring1][=Branch1]

# SELFIES to SMILES
benzene_smi = sf.decoder(benzene_sf)  
# C1=CC=CC=C1
\end{lstlisting}
In this example, benzene is first translated to \selfies and then back to \smiles. The initial \smiles string is dearomatized to encode the molecule robustly in \selfies.

\SkipTocEntry \subsubsection{Current capabilities of \selfies}
Currently, \selfies can represent ordinary organic molecules, including isotopes, charged and radical species. Furthermore it can represent chirality and stereochemistry by using an analogous approach to that of \smiles.

\selfies can not yet fully represent macromolecules, crystals, and molecules with \textit{complicated} bonds. We will explain the context, the challenges and potential ways to generalize \selfies to tackle these current shortcomings, and to develop an even more general, 100\% robust string representation for ML in chemistry.

\subsection{General Mappings}

\begin{figure}[htbp!]
\centering
\includegraphics[width=0.48\textwidth]{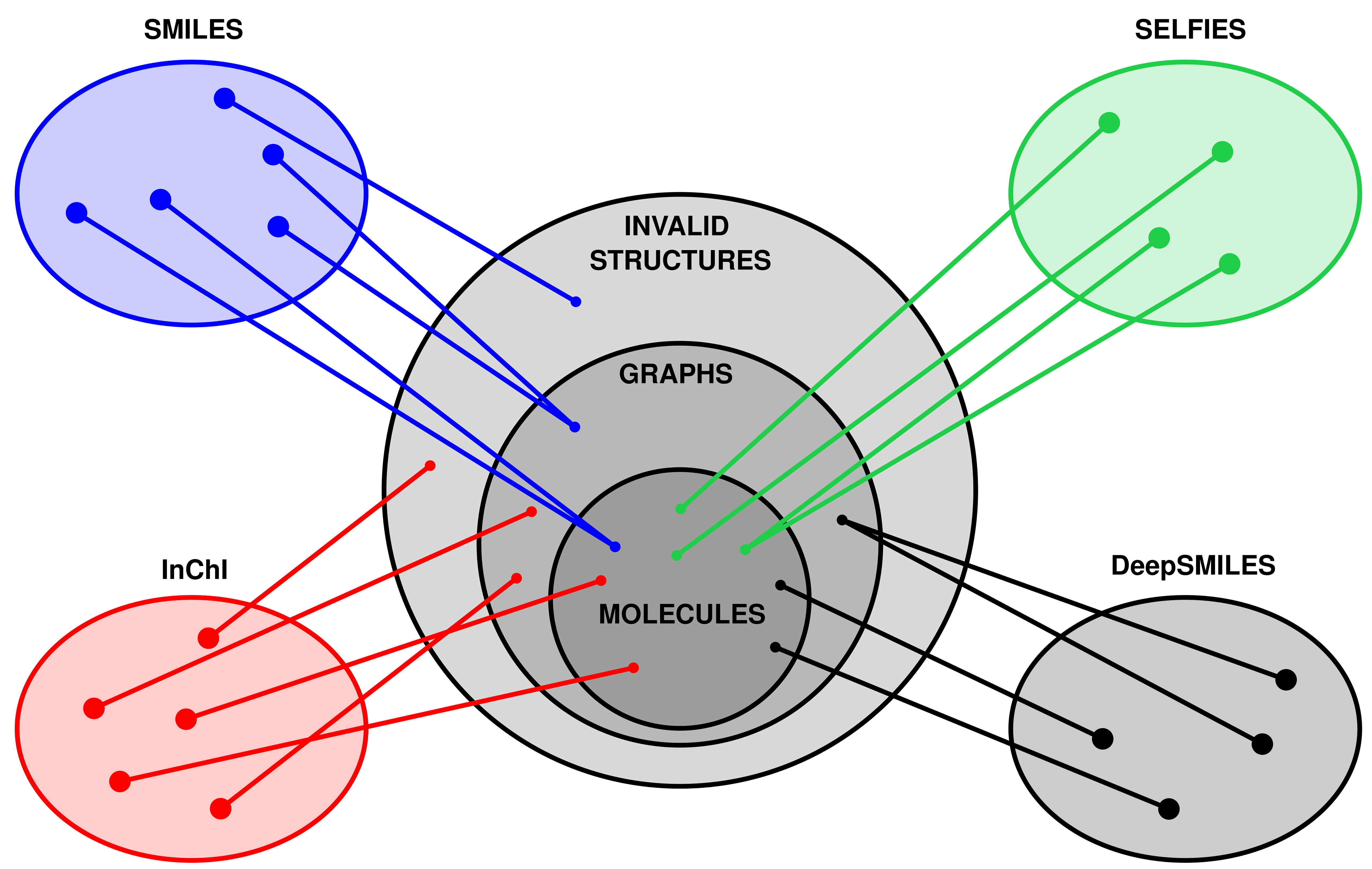}
\caption{Graphical representation of the mapping from strings to their corresponding structures. \smiles maps to general structures, that include molecules, but also non-molecular graphs or invalid (non-graph) structures. \inchi maps to the same space, although in a unique, bijective way. \deepsmiles maps strings to general graphs, not all of which stand for molecular graphs. Finally, \selfies is the only representation that maps in a surjective way only to molecular graphs.}
\label{fig:encodingSpaces}
\end{figure}

\selfies, \smiles, \inchi, and \deepsmiles are representations of a molecular graph. They all aim to map a string of tokens to a molecular graph, as illustrated in Figure~\ref{fig:encodingSpaces}. \smiles is a surjective representation from strings to structures that include molecular graphs, but also non-molecular (semantically invalid) graphs and other structures that cannot be interpreted as graphs (syntactically invalid). \inchi has the same co-domain, but its mapping is bijective, meaning each string corresponds to only one structure and vice versa. \deepsmiles makes the first important advance in terms of validity, and can be seen as a surjective mapping from strings to general (not necessarily molecular) graphs. Finally, \selfies is a surjective mapping from strings to molecular graphs. Both \smiles and \selfies can be made bijective through post-processing. For example, canonicalization (as provided by a number of tools such as RDKit) leads to a restricted domain, where each element maps to exactly one structure. It remains open whether a bijective mapping from strings to molecular graphs will be possible without post-selection. In the remaining text, we will discuss generalizations of \selfies and other molecular string representations along with important open questions. We will raise a number of concrete \textit{Future Projects}, which can be seen as stand-alone projects that aim to further the development of molecular string representation and their applications in ML for cheminformatics.

\stepcounter{TaskCounter}
\SkipTocEntry\subsection*{Future Project \arabic{TaskCounter}: meta\selfies~-- 100\% domain-agnostic robustness directly from data}\label{metaselfies}
So far, the discussion has focused on \selfies as a robust representation for molecular graphs. However, \selfies can also be thought of as a domain-independent robust representation for any graph in which vertices and edges have different semantic constraints. \selfies presently uses domain-dependent constraints which limit the maximum number of bonds which can be used by an atom. Mathematically, this constraint can be formulated in terms of the maximum vertex degree in a molecular graph. Interestingly, the domain-dependent rules could be obtained directly from large data sets in a deterministic way, without using ML. A technical description of such an algorithm is presented in the Supplementary Information of \cite{krenn2020self}. 

The derivation rules of \selfies are defined to satisfy the number of bonds a certain atom can form. In the language of graph theory, it constrains the vertex degree for each vertex type. Given a large enough data set of example graphs, one can directly approximate the maximum allowed vertex degree for every vertex type. Thus, \selfies obtains its defining feature of robust derivation rules.

It is important to realize that vertex degree constraints can not only be formulated for molecules in chemistry, but also for many other graph-based databases in the natural sciences. Examples include quantum optical experiments, where each individual optical element has a well-defined vertex degree constraint \cite{krenn2016automated}. In quantum circuits for quantum computers, individual gates have well-defined vertex degree constraints. RNA origamis \cite{han2017single} in biology also have vertex degree constraints (in addition to other constraints) that can be extracted from large databases.

Therefore, the robust generation of graphs can be seen as the basis of \selfies (meta\selfies) while the vertex degree constraints define the scientific domain. The opportunity of extracting the full \selfies language from data only and the understanding that this language can be applied in diverse domains opens up exciting opportunities. Given a particular data set, it would immediately, without training, be able to generate 100\% robust samples in the new domain, without anybody ever having to craft the language by hand. Additionally, a model could learn to solve design tasks in multiple domains. Given highly diverse training data sets, the opportunity for the generation of creative new solutions exists. For instance, one could us meta\selfies directly as the input of a Variational AutoEncoder (VAE) or a generative adversarial network (GAN). The quality of this approach will significantly depend on the size and diversity of the data set.

One can envision that domain-specific derivation rules could be shared in a standardized form in a \selfies registry, facilitating reuse by the community.

\stepcounter{TaskCounter}
\SkipTocEntry\subsection*{Future Project \arabic{TaskCounter}: The effect of token overloading in generative models}
One innovation in \selfies is the encoding of the sizes of branches and rings in a robust way. This is referred to as overloading and is done by enumerating the subsequent symbol(s) after the defining branch or ring token. Thereby, a token is interpreted as a hex number according to a table. A drawback of this way to ensure robustness is that it makes some \selfies more difficult to read. One important question is to understand how overloading impacts ML models, and whether the index alphabet -- which is currently heuristically composed -- can be improved to enhance performance in ML models. It might be interesting, using attention mechanisms, to study how these models understand overloading, and contrast with the way humans think about it.

\section{Macromolecules}

\begin{figure*}[htbp!]
\centering
\includegraphics[width=0.6\textwidth]{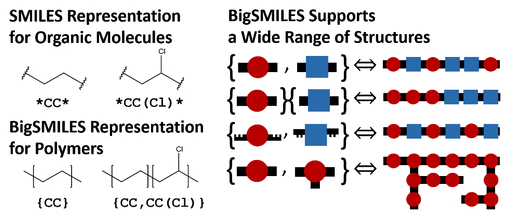}
\caption{Schematic of Big\smiles representations from Lin et al.\ \cite{lin2019bigsmiles} Polymers are represented as monomers (repeating units) enclosed within curly brackets; the curly brackets indicate that the molecule is a stochastic object. The monomers are represented as \smiles strings with additional information expressing the connectivity between monomeric units. } 
\label{fig:bigsmiles}
\end{figure*}
A challenging task in computational chemistry and biology is the simulation of macromolecules, which include biomolecules (nucleic acids, proteins, carbohydrates, and lipids) and synthetic polymers (e.g., plastics and synthetic fibers). Some macromolecules, such as polymers, are largely stochastic in nature and often feature a wide distribution over multiple chemical structures. In contrast, \smiles representations were created to describe deterministic structures such as small molecules, indicating that a new way of representing stochastic systems is needed.

One of the earliest macromolecule syntaxes developed was Curly\smiles \cite{drefahl2011curlysmiles}, which provides a method for encoding repetitive units such as monomers. This method encodes monomers as well-defined structures. Thus, it is unable to capture any stochasticity or complex connectivity between monomers. To address this issue, Lin et al.\ developed Big\smiles \cite{lin2019bigsmiles}, a polymer extension of \smiles which provides principles to represent the stochastic nature of polymers. A few syntax rules were added regarding the type of monomers and connectivity in the polymer. A schematic Big\smiles representations from Lin et al.\ is shown in Figure~\ref{fig:bigsmiles}. Big\smiles therefore provides a list of building blocks that can be assembled stochastically at run time. Since Big\smiles inherited the basic syntax of \smiles and introduced new symbols that require matching, it also suffers from the invalidity of some representations. 

Zhang et al.\ proposed \helm \cite{zhang2012helm} as a hierarchical way to represent large biomolecules. Unlike Big\smiles, which emphasizes the stochastic nature of synthetic polymers, \helm represents the full structure of a biomolecule with monomers replaced by their unique identifiers. This idea allows the representation of much larger structures in a concise way. \helm, however, has the same drawback as \smiles with respect to reliance on matching parentheses, leading to reduced robustness for its usage in generative models. 

Next we describe two interesting stand-alone projects that could advance molecular string representations and their application in AI for macromolecules.

\stepcounter{TaskCounter}
\SkipTocEntry\subsection*{Future Project \arabic{TaskCounter}: BigSELFIES~-- Stochastically assembling building blocks for 100\% robust polymers}

\selfies can naturally be extended to biomolecule representations by combining the best of Big\smiles (stochastic repeating patterns) and \helm (amino acids). A sequence of amino acids can be encoded with standardized symbols (for example, ``V'' = ``valine'') and every possible amino acid sequence is a valid representation. For the development of \helm-\selfies, one will need to identify grammatical rules for the entry and exit points of the amino acid sequence monomers or other macro-components. A challenge is that those rules likely go beyond individual bonding constraints, but this could be solved by adding more complex derivation states (i.e.,\ memory during the derivation).

From these rules, Big\selfies, an extension of \selfies to stochastic derivation using predefined lists of monomers, will follow directly. This is because \helm-\selfies will need to work for every combination of monomers. During derivation, it will not matter whether the structure is built deterministically or stochastically. 

Such a new representation will allow for the application of generative models to large molecules and polymers, with minimal hand-crafted features in the model. The ML algorithm can directly work on the string representation, and all outputs are valid and interpretable structures. This approach will allow for the applications of both simple and fast algorithms that have been proven successful for organic molecule design \cite{nigam2021beyond}. Furthermore, many deep generative models can directly be applied to design questions without any in-model conditioning or post-selection.

\section{Crystals}
A crystal is a periodic arrangement of atoms or molecules, commonly described by a set of lattice parameters, atomic coordinates, and symbols denoting symmetries other than translations. This description was standardized decades ago in the form of the Crystallographic Information File (CIF), which is widely accepted by the crystallography community \cite{hall1991crystallographic,brown2002cif}. The connectivity between atoms/building blocks is often a useful abstraction for thinking about chemical structures and materials which can be represented as a graph. The introduction of molecular graphs can be traced back to the 1870s \cite{professor1874lvii}, but it was not until the late 1970s that periodic graphs were introduced to describe crystals \cite{o1980plane,wells1977three}. Such abstractions led to various applications in solid-state chemistry. Prominent examples include the ``chemical diagrams'' used in the Cambridge Structural Database (CSD) for structure search \cite{groom2016cambridge}, connected coordination polyhedra to classify oxysalts \cite{krivovichev2009structural}, and net topologies in reticular chemistry \cite{o2008reticular}.

One can envisage an augmented version of \selfies that can be used to represent connectivity between atoms (the bond topology) in crystal structures robustly. String representations that have been explored for bond topology, such as the extended point symbols used in TOPOS \cite{blatov2014applied} for periodic graphs and the layered assemblies notation (LAN) \cite{tritsaris2020lan} for 2D materials, are either non-invertible (the graph cannot be constructed from strings without a lookup table), or based on a structural prototype. \selfies, however, provides a mapping that loses no information when converting between sequence and connectivity, and an explicit description of the connectivity. This allows for generative learning across the chemical space, and supervised learning on sequences instead of crystal structures or graphs. String-based graph representations are ubiquitous in chemistry and biophysics because strings are easy to use, process, and store, and there is a vibrant ecosystem of tools like RDKit and deep learning models for sequences that interface directly with strings. A robust string-based graph representation of crystals could inherit these advantages and transform materials informatics.

\begin{figure}[htbp!]
\centering
\includegraphics[width=0.44\textwidth]{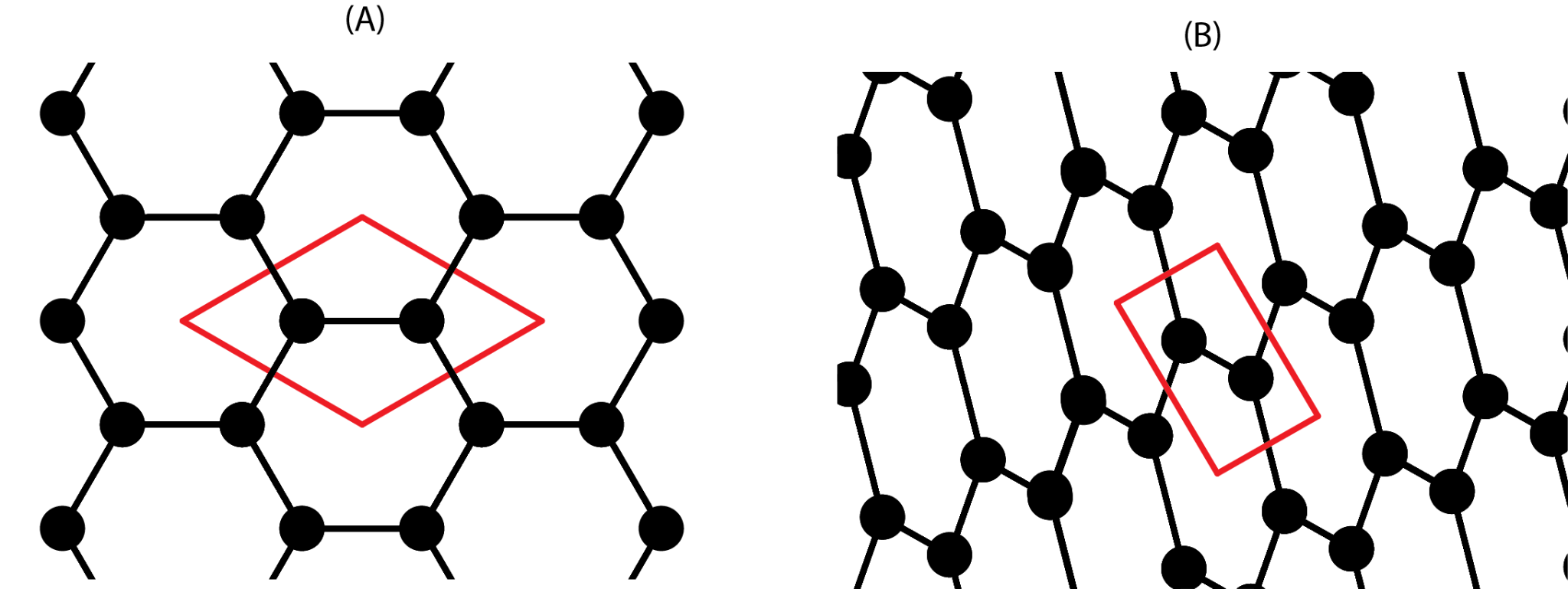}
\caption{Nets for representing crystals. (A) Crystal structure of graphene (2D honeycomb lattice). (B) 2D carbon structure of an orthorhombic lattice. The structures are two different faithful 2D embeddings of the same underlying net. This shows that a net, unlike its real space realization, does not bear spatial information (e.g., bond lengths, coordinates). Inspired by the success of \selfies in representing finite molecular graphs, in this section, we discuss how \selfies can be extended to represent crystal nets.}
\label{fig:crystals_fig0}
\end{figure}

\subsection{Net and quotient graph}
What is the ``crystal graph'' that can be represented by a string? To answer this question, first, the basic terminology used in this section is introduced. For more formal definitions, see Delgado-Friedrichs and O'Keeffe \cite{delgado2005crystal}. A crystal structure can be abstracted to a periodic graph, called a net, whose vertices represent the atoms (not atomic coordinates) and whose edges represent bonds between atoms. In practice, it might not be obvious which net best describes a crystal. The definition of edges can be ambiguous due to non-directional bonding or complicated coordination environments. For the latter, readers are referred to a recent benchmark of coordination number determination \cite{pan2021benchmarking}.

A net is an infinite, connected, undirected, simple (i.e., no loops and no multiple edges between a pair of vertices) graph. A net is $n$-periodic ($1\leq n \leq 3$) if it permits translations in $n$ independent directions. Assigning coordinates to vertices constructs an embedding of a net. An embedding is \textit{faithful} if edges do not intersect each other and only contain their respective end-vertices. Two faithful embeddings of the same net are shown in Figure~\ref{fig:crystals_fig0}. Note how they share the same net, even though they differ in their coordinates and cell parameters. Thus, to represent the connectivity in a crystal as a string requires representing a net that has a faithful embedding corresponding to the crystal's real space structure.

Generally, a graph with an infinite number of edges cannot be described by a string of finite length. Fortunately, a net can be represented by a finite graph, known as its \textit{quotient graph} \cite{chung1984nomenclature}. There are two variants of quotient graphs, one with directed, labeled edges, and one with undirected, unlabeled edges. Here, the focus will be on the former, which seems more suitable for developing crystal-\selfies (\textit{vide infra}) since only the first uniquely determines a net.

\begin{figure*}[htbp!]
\centering
\includegraphics[width=0.95\textwidth]{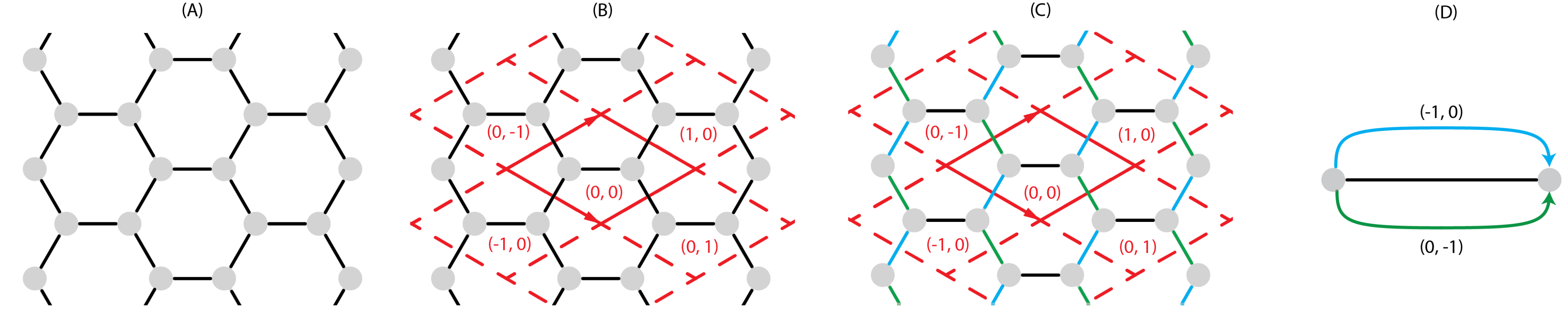}
\caption{Construction of the labeled quotient graph (LQG) for the underlying net of graphene. (A) Embed the net corresponding to graphene. (B) Define a coordinate system with two basis vectors (solid arrows) and an origin in the $(0,0)$ cell encompassed by solid lines. Index cells by their positions relative to the cell containing the origin. (C) Group bonds into three bond classes (black, blue, green) by translational invariance. (D) The result is the LQG. The label of $(0,0)$ bonds is dropped by convention.}
\label{fig:crystals_fig1}
\end{figure*}

The procedure to generate quotient graphs is depicted in Figure~\ref{fig:crystals_fig1} using graphene as an example: 
\begin{enumerate}
  \item Start from an embedding \textbf{E} of the net \textbf{N}.
  \item For embedding \textbf{E}, define a coordinate system \textbf{C} including an origin and a set of basis vectors (2 vectors for 2D, 3 vectors for 3D) representing the periodicity of \textbf{E}. Index all cells by their positions with respect to the origin. For instance, the cell containing the origin is the $(0,0)$ cell.
  \item Group translationally invariant edges into edge classes (black, green and blue in Figure~\ref{fig:crystals_fig1}).
  \item For each edge class, select one edge connecting a vertex in the $(0,0)$ cell and a vertex in the $(i,j)$ cell. Direct the edge starting from the vertex in the $(0,0)$ cell, and label this edge as $(i,j)$, where $i, j$ are restricted to \{-1, 0, 1\}. 
\end{enumerate}

The finite graph generated from this procedure is called the labeled quotient graph (LQG) of the embedding of the net \textbf{N} with coordinate system \textbf{C}. On the one hand, LQGs uniquely determine crystallographic nets up to isomorphism. An LQG can be converted to a net by choosing an arbitrary coordinate system, or to a crystallographic net through its automorphism group \cite{klee2004crystallographic}. On the other hand, LQGs with two different labelings can represent a pair of isomorphic nets. Such labelings are called equivalent. Methods to check for equivalent LQGs can be found in a study by Chung et al.\ \cite{chung1984nomenclature}.

An \textit{unlabeled quotient graph} UQG can be obtained by removing edge labels and edge directions from an LQG. UQGs are more similar to molecular graphs and preserve the neighborhoods of vertices. Unfortunately, the same (up to isomorphic) UQG could be derived from two nets that are not isomorphic, and \textit{vice versa} \cite{bader19973}. Thus, UQG alone cannot be used to describe a net. However, it is possible to enumerate LQGs from a UQG by enumerating edge labels \cite{thimm2004crystal}.

\stepcounter{TaskCounter}
\SkipTocEntry\subsection*{Future Project \arabic{TaskCounter}: Labeled Quotient Graphs in \selfies}
From the above definitions it appears that LQGs are most suited for string representation since they are 1) finite and 2) uniquely determine a net. LQGs have already been used in previous studies to represent crystals. A numerical encoding of LQG, the Systre key \cite{delgado2017crystal}, was implemented to identify nets. More recently, the LQG implementation was employed in crystal structure generation using a VAE \cite{xie2021crystal}. While the current \selfies scheme is able to represent molecules with localized bonds robustly, to represent an LQG, several improvements are needed:

\begin{enumerate}
  \item Edges in a quotient graph (LQG or UQG) can be self-loops or parallel edges; these are not allowed in the current \selfies. The solution may be to treat them as size 1 and size 2 rings, respectively.
  \item There should be symbols for edge directions and edge labels, such that the edge properties of an LQG can be represented.
  \item The choices for edge direction and edge label are finite, and not all labelings are allowed, for example, parallel edges cannot have the same labeling vector $(i,j)$. There should be additional grammar that respects such (often local) restrictions.
  \item While an LQG uniquely determines a net, two non-isomorphic LQGs can represent the same net. This can happen in many cases, such as constructing an LQG from a supercell or from the aforementioned label equivalence. Thus, a canonicalization process is desired such that every net can have a canonical crystal-\selfies.
\end{enumerate}

\stepcounter{TaskCounter}
\SkipTocEntry\subsection*{Future Project \arabic{TaskCounter}: Crystal-\selfies in generative models}
The search space for theoretical materials is practically infinite. While high-throughput virtual screening methods are now common in materials informatics and valuable for exploring new regions of materials space, generative models could provide a more systematic direction for targeted materials design. Generative models also aim to reduce systematic bias in the exploration of chemical space, allowing for a higher chance of discovery. By solving the missing pieces in the previous \textit{Future Project}, \selfies could be augmented to crystal-\selfies, a lightweight and robust string representation of crystal (bond) topology that could improve crystal structure generation.

Currently, a few different approaches are followed to construct generative models for crystal structures. The first approach, employed mainly in the field of metal-organic frameworks (MOFs) \cite{yao2021inverse,colon2017topologically}, starts from a net that is usually selected from established data sets. Appropriate building blocks are then chosen as nodes and their connections as edges of the net. The generation resembles the isoreticular expansion of MOFs. Such a method relies on predefined nets in addition to a set of available building blocks.

Another approach is to focus solely on embeddings. The embedding can be represented by a set of parameters based on a structural prototype \cite{fung2021inverse,nouira2018crystalgan}, which may not be generalizable. Alternatively, embedding representations can be learned \cite{court20203,noh2019inverse} from data sets. Such representations are often continuous, thus suitable for inverse design. However, since bond topology information is not explicitly included, it is unclear whether this approach can generate topologically diverse structures.

Alternatively, it is possible to start with generating LQGs: In 2004, Thimm demonstrated that structures can be generated with minimal specifications (number of atoms in a unit cell and vertex degree for each atom) by 1) generating a UQG based on the specifications, 2) enumerating LQGs from the UQG, 3) unfolding the LQGs to nets, and 4) obtaining faithful embeddings from the nets \cite{thimm2004crystal}. This method allows to control the formation of types of nets over generated structures and does not rely on predefined nets. In addition, as discussed earlier, both LQGs and UQGs can be represented by crystal-\selfies. Thus, following Thimm's approach, structure generation using crystal-\selfies can be, for example, a mapping: chemical composition $\to$ UQG (crystal-\selfies) $\to$ LQG (crystal-\selfies) $\to$ net $\to$ embedding.

A shortcoming of net-based representations is the obscure connections between the net of a crystal and the physical/chemical properties of that crystal. From a \smiles string or a molecular graph, properties (e.g., 2D descriptors) like log\textit{P} can be readily estimated without embedding the graph (i.e., molecular conformations). However, for crystals, currently, both physical and chemical properties are calculated from embeddings. Thus a calculator connecting net and crystal properties would greatly benefit the development of this field. It has been demonstrated that the dimensionality of a crystal structure can be derived from its LQG \cite{gao_determining_2020}. More information regarding relations between a net and its embeddings can be found in a study by Blatov and Proserpio \cite{blatov2010periodic}.

Finally, for crystal generative models using \selfies, some general considerations are listed here:
\begin{enumerate}
  \item The alphabet of \selfies can be extended to include building units and linkers used in reticular or inorganic chemistry. This also helps to minimize the space of LQGs by reducing the number of vertices. An alternative would be to use contraction operations.
  
  \item It has been demonstrated that the symmetry and topological features of an LQG are related to that of the corresponding net \cite{thimm2009crystal,eon2016topological}. Thus, the model can be conditioned on these features.
  \item While a UQG does not determine a net, it does preserve neighborhoods. This means that it is possible to generate nets with specific local structures by making the neighbors of a vertex immutable. 
  \item Some nets cannot be (faithfully) embedded in 3D. Crystal generative models should be conditioned such that these ``pathological nets" are excluded from generations. Some properties used to identify such nets are introduced by Thimm \cite{thimm2004crystal}.

\end{enumerate}

\section{Beyond organic chemistry: Complicated bonds}

\begin{figure*}[htbp!]
    \centering
    \includegraphics[width=0.75\textwidth]{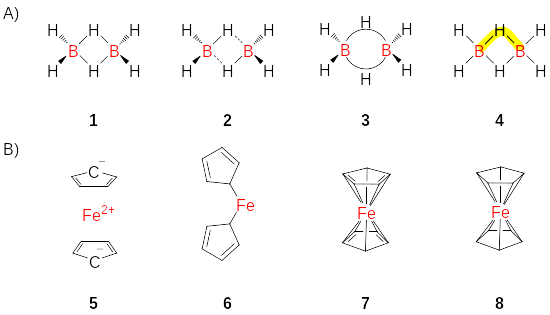}
    \caption{A) Different structural representations for diborane (B$_2$H$_6$), where \textbf{1} properly accounts for the symmetrical B$_2$H$_6$ ``diamond core'' but gives an incorrect valence electron (VE) count, \textbf{2} uses ``zero-order bonds'' indicated as dashed lines to preserve the VE count but features a molecular symmetry that is too low, \textbf{3} attempts to capture the actual three-center two-electron (3c-2e) bonding by use of arced ``banana bonds'' but cannot be used in molecular graph approaches which only allow for each edge to connect two nodes (atoms), and \textbf{4} shows the full delocalization of an electron pair over the B-H-B unit.
    (B) Lewis structures of ferrocene (C$_{10}$H$_{10}$Fe), where \textbf{5} is unfortunately used by PubChem, but is wrong as the compound is not ionic. \textbf{6} and \textbf{7} cannot account for the $^1$H and $^{13}$C NMR spectra, both of which feature only one singlet indicative of ten chemically equivalent CH units. Only \textbf{8} is fully in line with crystallographic and spectroscopic data, but at the expense of making electron counting impossible.}
    \label{fig:sec6_diborane_ferrocene}
\end{figure*}

In this chapter, we discuss the challenges and prospects of extending \selfies beyond organic chemistry. In contrast to organic molecules \cite{goodman2021inchi}, transition metal, lanthanide, actinide, and main-group metal compounds are difficult to handle with current digital molecular representations \cite{warr2011representation} due to special bonding situations and intricate 3D structures, combined with technical limitations that have evolved for historical reasons. Most problems trace back to 1) the assumption that bonding is localized and thus can be described with valence bond (VB) theory, 2) the non-explicit representation of terminal hydrogen atoms, which are added to the heavy (non-H) atoms based on rules derived from VB models in an approach called ``implicit hydrogens'', and 3) the inability to describe stereochemistry that goes beyond the usual restrictions of organic chemistry, i.e.,\ stereogenic carbon centers plus some cases of \textit{cis}/\textit{trans} isomerism in C=C double bonds and cumulenes. While organic chemistry has plenty of examples of more advanced stereochemistry such as planar and helical chirality \cite{Pfaltz5723,narcis2014helical,lopez2022planar}, current digital molecular representations are generally not equipped to handle those.

Therefore, any approach toward a general digital molecular representation covering all elements of the periodic table will fail if it is unable to handle the issues mentioned above. Here, we will illustrate a number of prominent examples which highlight the urgent need to improve the situation, as otherwise a major part of chemical space will remain inaccessible to modern cheminformatics and AI approaches \cite{huang2021ab}.

\subsection{Complex, ``fuzzy'' bonding situations \textit{vs}.\ valence bond theory}
One reason for including connectivity information in a molecular string representation is that it allows chemists to describe structures in a simple way, for example, by decomposing them into substructures. Furthermore, from an ML perspective, connectivity information might also be thought of as an additional inductive bias that can help a model to generalize \cite{wilson2020bayesian}.

However, bonding information turns into a significant technical problem if there is no algorithmically unambiguous way to define it \cite{gonthier2012quantification}, and when there is a wide array of possible interactions of different strength and origin. This ambiguity in defining bonds has led some chemists to call them ``convenient fiction'' \cite{ball2011beyond}, which is also reflected in the widespread use of the bond type ``Any'' for substructure queries in databases such as the CSD, to ensure no entries are missed. In some domains of chemistry, VB theory provides a convenient and intuitive way to think about chemical bonding that is easy to encode in widely used data structures. In standard organic chemistry, for instance, most bonding situations can be described as two-center two-electron (2c-2e) bonds, a scenario which translates well into molecular string representations where atoms are nodes and covalent bonds between two atoms sharing two electrons are edges of a molecule graph. However, as the Open\smiles standard notes ``This simple mental model has little resemblance to the underlying quantum mechanical reality of electrons, protons, and neutrons\ldots'' \cite{james2015opensmiles}.

Two prominent examples from main group element and transition metal compounds, respectively, will be discussed here to outline the corresponding major issues. Figure~\ref{fig:sec6_diborane_ferrocene}A shows four different molecular structural models for diborane (B$_2$H$_6$), an important reducing agent and key reactant for hydroboration reactions. Most (inorganic) chemists, when asked to sketch the molecule, will likely draw structure \textbf{1}, which properly captures the two bridging $\mu_2$-hydrido ligands, but results in an incorrect valence electron (VE) count of 16 VEs instead of the proper 12 VEs, when each line connecting two element symbols is assumed to represent two electrons. In order to preserve the electron-counting function of the lines representing 2c-2e bonds, sometimes structure \textbf{2} is used, wherein additional interactions between the two BH$_3$ subunits are indicated by dashed lines, which are assumed not to contribute to the electron counting, and thus have been termed ``zero-order bonds'' by Clark \cite{clark2011accurate}. However, this structure \textbf{2} incorrectly implies the symmetry of the molecule to be C$_{2h}$, while X-ray structure analysis has demonstrated that diborane belongs to the D$_{2h}$ point group. All four terminal B--H bonds are equivalent at approximately \SI{1.09}{\angstrom}, and the four B-H distances in the B$_2$H$_6$ ``diamond-shaped core'' are also essentially equivalent at about \SI{1.24}{\angstrom}. Notably, the observed differences of $<$\SI{0.03}{\angstrom} in these formally equivalent B--H bond distances are possibly caused by packing effects \cite{smith1965single}. Therefore, some chemistry textbooks use structure \textbf{3} with two bent ``banana bonds,'' with the two arched lines each representing two VEs. Such a representation, although it gives the correct VE count, cannot be used in standard molecular graphs, which assume that each edge connects two -- and only two -- nodes (atoms). A better description of the structure of diborane makes use of three-center two-electron (3c-2e) bonds, where two electrons are fully delocalized over the B-H-B unit, as highlighted in yellow in structure \textbf{4}.

Another complex bonding situation arises in organometallic ``sandwich'' complexes such as ferrocene (C$_{10}$H$_{10}$Fe), which are common building blocks in organic chemistry and have important industrial applications, for example in Ziegler-Natta catalysis \cite{janiak1998metallocenes}. Some databases such as PubChem \cite{kim2021pubchem} utilize ionic structure \textbf{5}, as shown in Figure~\ref{fig:sec6_diborane_ferrocene}B, assuming a ``naked'' Fe(II) cation without any coordinated ligands, combined with two separate cyclopentadienyl anions. This structure, however, is utterly wrong, as ferrocene is a compound without separate charged ions, that can be purified by vacuum sublimation and is non-soluble in polar solvents such as water, but dissolves well in non-polar organic solvents such as \textit{n}-hexane and toluene. The uncharged structure \textbf{6} would be in line with these properties, but does not account for the $^1$H and $^{13}$C NMR spectra, which both exhibit only one single peak, indicating that all ten CH units are chemically equivalent, while the NMR spectra of representation \textbf{6} would feature three different peaks for each nucleus. Furthermore, two-coordinate iron centers are exceedingly rare and require very bulky ligands to be stabilize \cite{sharpe2018selective}. Alternatively, structure \textbf{7} has the Fe(II) center ``sandwiched'' between the two cyclopentadienyl rings, but still cannot account for the NMR spectra due to the combination of two localized C=C double bonds and one carbanionic center per ring. Only structure \textbf{8} correctly captures both the NMR properties and the X-ray data, which indicate ten equivalent Fe--C and C--H bonds, and an identical length for all ten C--C bonds \cite{dunitz1956crystal}. This, however, goes at the expense of any kind of VE counting, as the actual bonding requires a molecular orbital (MO) treatment which at least considers both the cyclopentadienyl $\pi$-system and the iron \textit{d} orbitals.
The situation becomes even more complicated when one attempts to capture not only covalent bonds but also weaker agostic interactions, in which the two electrons of a C--H bond interact with empty metal \textit{d} orbitals in another example of 3c-2e bonding. The same applies to other weak interactions such as hydrogen bonds, raising important questions as to which interactions should actually be captured in a digital molecular representation as a ``bond'' (and which not), and how to automatically detect them from a set of atomic coordinates, ultimately leading to a rather arbitrary distinction between \textit{bonded} and \textit{non-bonded}. To quote Democritus: ``Nothing exists except atoms and empty space; everything else is only opinion.''

\begin{figure*}[htbp!]
    \centering
    \includegraphics[width=0.7\textwidth]{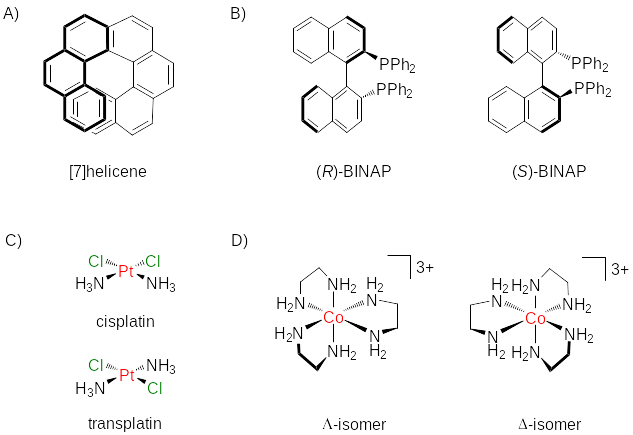}
    \caption{Examples of (A) helical and (B) axial chirality in organic compounds. (C) Diastereomeric coordination compounds: cisplatin is an approved anticancer drug, while its isomer transplatin is inactive. (D) Helical chirality in metal complexes.}
    \label{fig:sec6_chirality}
\end{figure*}

\subsection{No ``standard" valences}
Current molecular string representations make use of models based in VB theory, as it allows the definition of ``standard valences'' for the different elements. Missing hydrogen atoms are inferred and inserted implicitly, which allows for a more compact representation. These standard valences are usually fixed to satisfy the octet rule, which is not generally applicable. Even many main group elements do not follow that rule. For the \textit{d} and \textit{f} elements, such a rule is largely irrelevant due to strongly delocalized bonding with significant mixing between metal and ligand orbitals that require an MO theory treatment, something that cannot be captured by structural representations exclusively based on 2c-2e bonds.

For example, while the noble gas elements have to be formally assigned a ``standard valence'' of zero, many stable compounds with them, such as XeOF$_4$, are known and readily prepared. Even carbon does not necessarily obey the octet rule, as the catalytic center of nitrogenase, the enzyme that is central for biological nitrogen fixation, contains an FeMo co-factor with the composition of [Fe$_7$MoS$_9$C] that is built around a carbide center with a formal charge of $-$IV and six equivalent Fe--C bonds, as demonstrated by X-ray crystallography \cite{einsle2020structural}. Beyond such surprising structural motifs created by nature itself, inorganic chemists in particular constantly look for new oxidation states \cite{yu2016oxidation} and bond orders \cite{la2008bond, nguyen2005synthesis}. Furthermore, there needs to be a critical discussion of the term ``valence'' itself, as in inorganic chemistry it is normally used to describe the physical oxidation state (related to the spectroscopically accessible \textit{d}-electron count) of a metal center (e.g., trivalent iron is Fe(III), which is usually six-coordinate), while in the context of \inchi and \smiles, it refers to the number of bonds to neighboring atoms. Therefore, any approach to generally applicable digital molecular representations should not make use of ``standard valences'' and needs to treat all hydrogen atoms explicitly. 

\subsection{Stereochemistry beyond the tetrahedron}
Most organic molecules feature either linear sp, planar sp$^2$ or tetrahedral sp$^3$ carbon centers and, thus, their stereochemistry is usually restricted to point chirality from stereogenic centres, \textit{cis}/\textit{trans} isomerism of C=C double bonds in alkenes or axial chirality in allenes/cumulenes. However, in more complex structures, even within organic chemistry, planar or axial chiral elements can additionally come into play. Prominent examples of the latter include \textit{ortho}-condensed polycyclic aromatic compounds from the class of the [\textit{n}]helicenes (Figure~\ref{fig:sec6_chirality}A). Such systems are far from academic curiosities, as axial chirality is important to enantioselective catalysis. This is apparent in the BINAP class of ligands, for which Noyori was awarded the 2001 Nobel Prize in Chemistry (Figure~\ref{fig:sec6_chirality}B).

Furthermore, metal complexes are characterized by a wide range of coordination geometries with coordination numbers in the range of 2--16. The structural motif assumed is often dictated by electronic ligand field (LF) effects rather than steric repulsion, as in the widely used VSEPR model applicable to main group chemistry. For example, a metal center with four ligands, in addition to a tetrahedral structure, could also assume a square-planar coordination environment, where the central metal atom and the ligands are in one plane, with L-M-L angles of 90$^o$ and 180$^o$, respectively. In MA$_2$B$_2$-type compounds, this gives rise to two stereoisomers, with \textit{cis}- and \textit{trans}-[PtCl$_2$(NH$_3$)$_2$] as some of the most important examples (Figure~\ref{fig:sec6_chirality}C). The compound cisplatin is an approved anticancer drug with wide applications in chemotherapy and annual multi-billion dollar sales, while transplatin shows no biological activity. Unfortunately, PubChem considers both compounds simply as ``synonyms'' and thus provides an incorrect record for them \cite{national2021pubchem}. The reason for this is rooted in the erroneous application of the concept of ``standard valences.'' Since the Pt(II) center is assigned a valence of two, the compound is incorrectly represented as a mixture of a bent(!)\ PtCl$_2$ unit and two separate NH$_3$ molecules to also preserve the ``standard valence'' of three for nitrogen. However, the two ammine ligands are bonded to the metal in a fashion that is comparable to covalent bonds in organic chemistry, and in aqueous solution it is actually the chlorido ligands which are exchangeable to water, not the ammine ligands. When moving from four- to six-coordination, the range of accessible structures becomes even broader and one has to additionally consider new stereocenters generated by fixation of ligand atoms to the metal, which can lead to helical structures, as discovered by Alfred Werner more than 100 years ago \cite{werner1913constitution} (Figure~\ref{fig:sec6_chirality}D). To complicate matters even further, coordination numbers of 12 and higher have been reported. One example is [Ph$_4$P][Hf(BH$_4$)$_5$], in which each borohydride unit [BH$_4$]$^-$ acts as a tridentate ligand to the Hf(IV) metal center, which has a coordination number of $5 \times 3 = 15$ \cite{makhaev1990anionic}.

\subsection{Alternative approaches}

\begin{figure*}[htbp!]
    \centering
    \includegraphics[width=0.9\textwidth]{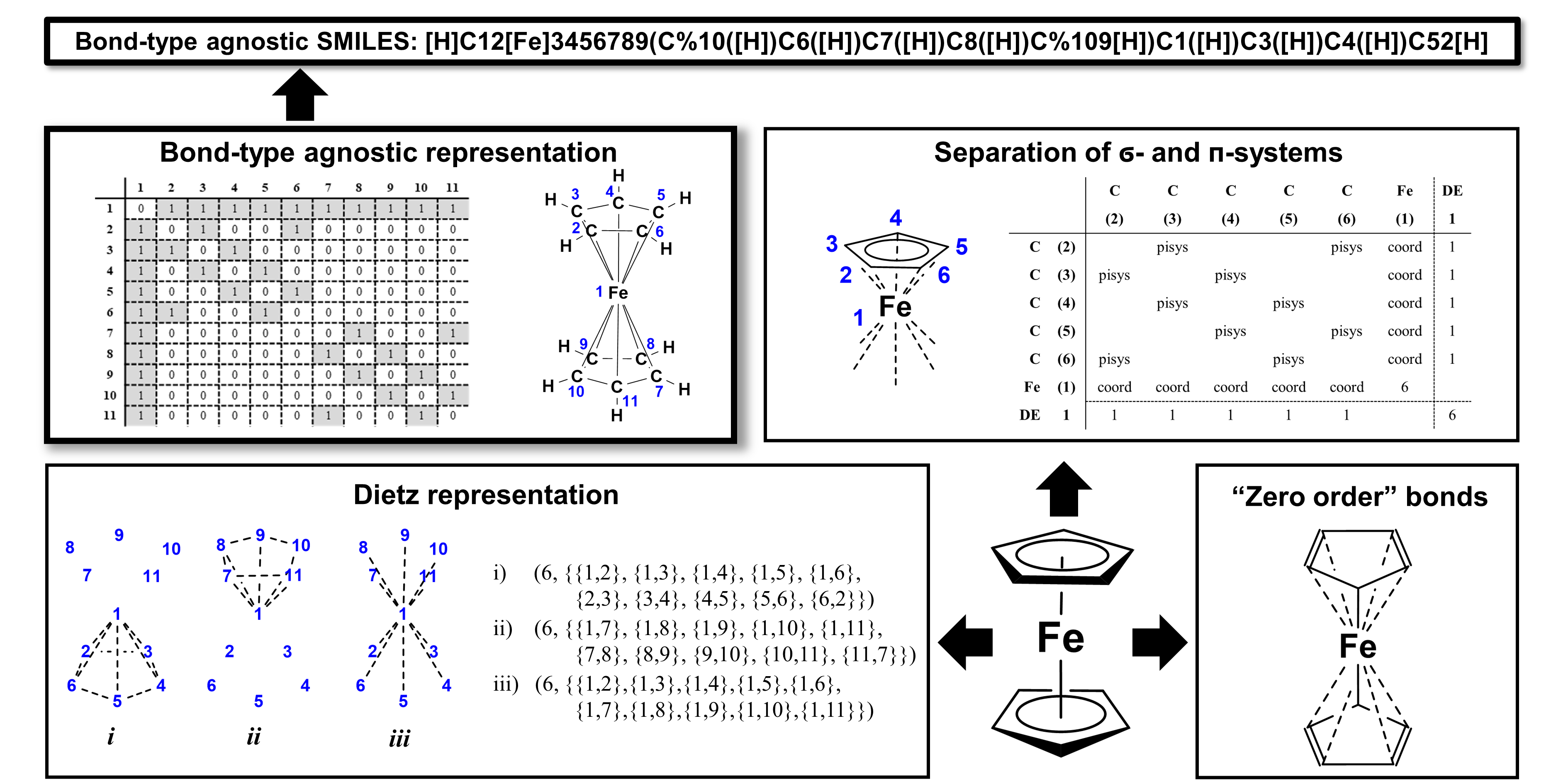}
    \caption{Current possibilities to represent molecules with complicated bonds (here ferrocene). Top Left: Bond-agnostic edges neglect some physical constraints, and can be written as \smiles or a graph. Top Right: Separation of $\sigma$- and $\pi$-electron systems. Bottom Left: Dietz representation. Bottom Right: ``Zero-order" bonds.}
    \label{fig:sec7_complicated_bonds}
\end{figure*}

Many alternative molecular representations that have been put forward try to be more faithful in representing chemical concepts such as multicenter bonds or stereochemistry.

\SkipTocEntry\subsubsection{Separation of \texorpdfstring{$\sigma$}$- and \texorpdfstring{$\pi$}$-electron systems}
 
In conventional molecular string representations (e.g., \smiles and \selfies), atoms are considered to be nodes and bonds to be edges of a molecular graph. These are then assigned numerical values such as atomic number, number of unshared electrons, and bond order, which are considered invariants of the graph, as they do not depend on the labeling scheme of the nodes (atoms) \cite{krotko2020atomic}. Most approaches allow all edges to connect just two nodes, in line with the standard 2c-2e bonds that dominate most of organic chemistry. 

In the symbolically extended BE (sXBE) matrices \cite{ugi1971beschreibung, ugi1993new, stein1995new, DBLP:phd/dnb/Stein93}, however, delocalized electron systems are encoded using special bond types such as \texttt{pisys} (e.g., benzene) or \texttt{edsys} (for electron deficient systems such as boranes). Therefore, these representations allow for a better representation of the true multicenter bonding nature of some systems such as diborane or ferrocene (Figure~\ref{fig:sec7_complicated_bonds}A--B).


\SkipTocEntry\subsubsection{Dietz representation}
As an alternative, Dietz suggested a hypergraph concept \cite{dietz1995yet} where edges are allowed to contain more than two nodes, accounting for multicenter bonding (Figure~\ref{fig:sec7_complicated_bonds}C). However, the approach of Dietz, Ugi, and Stein is based on groups of nodes and edges which are additionally characterized by the number of unshared VEs and delocalized electrons \cite{krotko2020atomic}. This approach tries to exactly capture the electronic structure but leads to complicated nested sets of brackets that may be hard to comprehend. Furthermore, a clear assignment of valence electrons is often not possible transition metal chemistry due to extensive delocalization. Consequently, as the resulting representation and terminology is difficult to tackle, to our knowledge, they have not been used in any digital structure representation to date. Furthermore, as noted by Bauerschmidt and Gasteiger, the Dietz system (and all others described so far) cannot easily distinguish between different spin states of the electrons \cite{bauerschmidt1997overcoming}. This is relevant for carbenes, where the singlet and triplet states have a vastly different reactivity, and also applies to molecules as simple as dioxygen. Hence, together with its complexity, this representation has not found widespread use.

\SkipTocEntry\subsubsection{``Zero-order" bonds}
To address the issue of multicenter bonding, non-specified bond orders, and the related problems with implicit hydrogens, in 2011, Clark proposed two backward-compatible modifications to connection table (CT)-based molecular representations \cite{clark2011accurate}. In that work, it was suggested to allow for a bond order of zero for all interactions or bonds that do not fit the conventional scheme and to add a property that explicitly describes the number of connected hydrogen  (Figure~\ref{fig:sec7_complicated_bonds}D). Interestingly, the zero bond order reflects the fact that, due to the ambiguity of bond orders, many chemists perform database substructure searches with ``Any'' as bond type. However, as discussed in the previous section (Figure~\ref{fig:sec6_diborane_ferrocene}A, Structure \textbf{2}), this can lead to an incorrect decrease in molecular symmetry. There are also cases where ambiguities appear regarding which bonds should be denoted as ``zero-order'' and which ones otherwise. A common resort to be expected in that context is that many users will then simply label all bonds as ``zero order.''

Thus, it should be stressed again that in \textit{d}- and \textit{f}-block chemistry as well as main group organometallic compounds it is often impossible to assign any particular bond orders without high-level quantum chemical calculations, due to the highly delocalized nature of the bonding, where electrons are often spread out over a significant number of atoms, including the metal center itself, the immediately coordinated atoms, and additional ligand groups. In summary, despite more than 25 years of research into the issue, little progress has been made toward a generally applicable and domain-independent digital molecular representation, as some of the concepts representations are built upon (standard valences, 2c-2e bonding, and the possibility to assign bonds and bond orders unambiguously) are ill-defined for many compounds outside of classic organic chemistry.

\SkipTocEntry\subsubsection{Tooling and the value of simplicity} 
In this section, a number of essentials characterizing molecular assemblies of atoms and what is needed to create a digital representation thereof are outlined. The high variability of metal complexes, in particular in terms of electronic structure and coordination geometry calls for a flexible and extensible ``layer approach'', in which the essentials strictly required to describe a molecular structure are included in a \textit{base layer} while all domain-specific information is covered by additional and user-definable \textit{property layers}, which can be used or ignored depending on the users' goals. 

\begin{enumerate}
  \item \textbf{Base layer (domain-independent)}: The nodes (atoms) ``carry'' the atomic number and (non-standard) isotope distribution. Edges (bonds) indicate strong pair-wise attractive interactions, although it remains to be defined which interactions should be captured and which ones not.
  \item \textbf{Property layer \#1 (domain-dependent)}: Nodes carry information about local stereochemistry and charge; edges carry bond order and type information (such as single, double, triple, aromatic bonds).
  \item \textbf{Higher-level property layer \#2 (domain-dependent)}: Information from ML models, handcrafted information, experimental data such as NMR chemical shifts, ``strategic bonds'' for either retrosynthesis or reactivity prediction.
\end{enumerate}



An interesting aspect of the additional \textit{property layers} is that, beyond certain values assigned based on user interaction or software-encoded domain-specific models, these might also be generated from ML approaches, which could allow for a more nuanced picture than simple binary assignments often governing current models \cite{jablonka2021using}. To conclude, the need to describe all of chemical space is at odds with imposing strong rules on the allowed valence or connectivity and more elaborate derivation rules need to be developed.

\stepcounter{TaskCounter}
\SkipTocEntry\subsection*{Future Project \arabic{TaskCounter}: Automatic compilation of complex rules from data}

Most chemists may possibly agree on which structures are ``correct'' (and which are not) by visual inspection of structural formulas. As this ability is based on knowledge obtained by inspection of other compounds and the underlying trends that govern their bonding, it should be possible to train an ML model to deduce these rules (i.e.,\ the necessary extended \selfies grammatical rules) for general \selfies from an appropriate data set. This project is a further extension of the topic described in \textit{Future Project} 1 (meta\selfies). One of the most extensive and curated structure collections is the CSD. However, one has to keep in mind that there will be biases in such a data set which need to be accounted for. For example, the CSD only contains compounds that could be crystallized and were deemed to be of sufficient interest for X-ray structure analysis. This could potentially be corrected by supplementing the model with data from other databases and by the addition of manually selected structures. Furthermore, state-of-the-art quantum chemical calculations are nowadays able to provide optimized geometries that often approach the accuracy of experimentally obtained structures and might thus also be of interest to feed to such models. One potential means of progression is to create a neural network that learns to classify compounds into ``correct'' or ``incorrect'' categories. After training, symbolic regression \cite{cranmer2020discovering} could be used to extract symbolic rules that can be used directly by \selfies.

\section{Reactions}
\begin{figure*}[htbp!]
    \centering
    \includegraphics[width=0.95\textwidth]{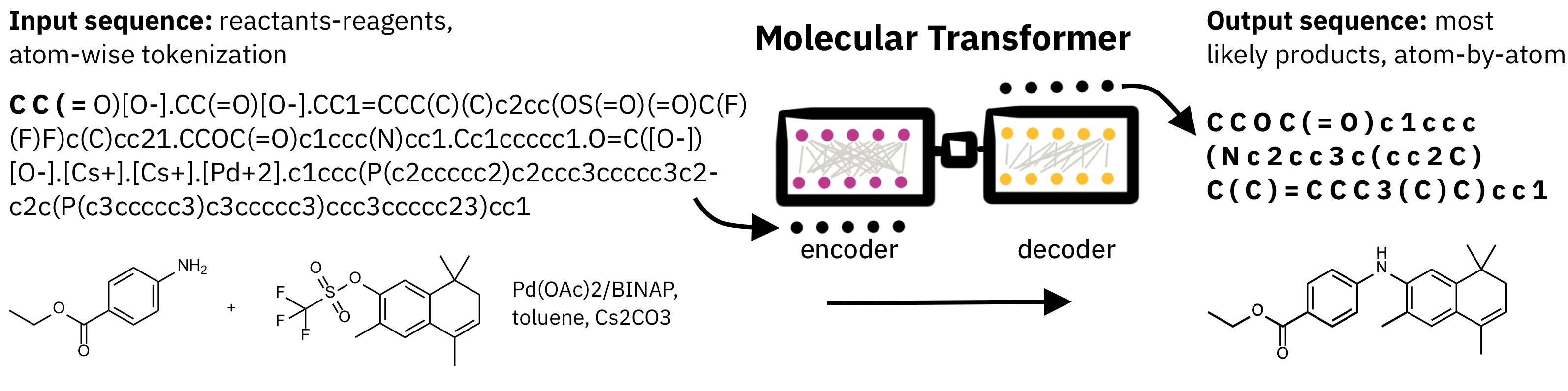}
    \caption{An example of a Molecular Transformer which uses \smiles to represent and transform reactant and agent molecules into the product of the reaction, as used by Schwaller et al.\ \cite{schwaller2019molecular}. The tokenization of the \smiles is shown by the bold characters separated with spaces.}
    \label{fig:molTransformer}
\end{figure*}

So far, we have discussed only representations of molecules. However, a significant part of chemistry consists of the modifications of molecules, via reactions. In this section the applications of ML in reactions is discussed and what role molecular representations play.

A chemical reaction can be divided into four distinct parts: reactants, agents, products, and overall conditions. Products are the outcome of the reaction, or the molecule(s) obtained once the reaction is done. Reactants are the building blocks of the product(s): the initial compounds containing atoms that will be incorporated into the product. Agents can be anything from catalysts to solvents that are added to the reaction mixture but will not be part of the product molecule(s)\footnote{This is a simplification as sometimes it is not possible to identify which molecule contributes to the product, such as in reactions involving protic catalysts.}. Conditions are, for example, the temperature and pressure at which the reaction is run, or other more complex variables such as heating profiles, the order of addition of reactants and agents, and so on. The agents and conditions describe the environment in which the reaction happens. Depending on the available data set, conditions and agents may not always be fully described.

Openly available data sets are derived from either patents \cite{Lowe2012,Lowe2017} or chemical journals \cite{jiang2021smiles}, and more rarely experimental procedures directly \cite{SantanillaScience}. These data sets are distributed using \smiles as a representation for the reaction itself and usually include extra information in various formats. There is no standard format that allows for conveying information about reactions and their details simultaneously. Initially intended for organic chemists, these data sets also attracted the attention of computational chemists as they enabled the development of new methods and algorithms. The Open Reaction Database provides a centralized platform to collect and access reaction data sets \cite{kearnes2021open}.

Chemical reactions are commonly investigated in ML for chemistry regarding two broad categories: reaction completion and property prediction. Usually, the full reaction is provided when running property predictions. A typical variable to predict could be the yield of the reaction or the energy profile. Reaction completion consists of completing a reaction scheme, where some of the molecules or conditions are missing. Two subcategories of interest are reaction prediction, where the goal is to predict a product based on a given set of reactants, and retrosynthesis, where the goal is to predict a set of reactants given a particular product. Likewise, prediction of reaction conditions and/or agents represents a major current challenge.

\begin{enumerate}
  \item Reaction completion
  \begin{enumerate}
    \item Reaction prediction
    \item Retrosynthesis
    \item Condition and agent prediction
  \end{enumerate}
  \item Property prediction
\end{enumerate}

Reaction completion is the category of tasks where the representation matters most, as algorithms not only take molecules as input but also need to output molecules. Therefore, the main discussion here will be about possible algorithms and representations of reactions with respect to reaction completion.

There are three broad categories of methods designed for reaction completion:
\begin{enumerate}
  \item Template-based methods
  \item Graph-based methods
  \item Text-based methods.
\end{enumerate}

Template-based methods use a set of reaction templates that encode the possible changes effected during a reaction. These templates are either written by domain experts \cite{szymkuc_computer-assisted_2016} or directly extracted from data using atom-mapping \cite{wcoley_graph-convolutional_2019, segler_planning_2018}. Atom-mapping links the product atoms with the corresponding reactant atoms, and, hence, specifies the reaction center. In template-based reaction completion methods, it is common to see the outcome of these templates ranked by a neural network \cite{wcoley_graph-convolutional_2019, segler_planning_2018} to define which reaction is the most likely to happen. Graph-based methods \cite{wcoley_graph-convolutional_2019, jin_predicting_2017} typically use graph neural networks (GNNs). Generally, this kind of method splits the project into two sub-tasks: the first step localizes where the changes in the graph should happen by selecting atoms, and in a later step, the changes are performed. Similar to template-based methods, the bond changes used for training of the graph-based methods are extracted from atom-mapping. Therefore, their performance depends on the quality of the underlying atom-mapping \cite{schwaller2021extraction}. 

Text-based methods use textual representations of molecules to take advantage of models initially developed for neural machine translation, such as the Transformer model \cite{vaswani_attention_2017} (see Figure~\ref{fig:molTransformer}). Such sequence-2-sequence methods for forward prediction, retrosynthesis, and agent completion can be atom-mapping independent as the reactant and product atoms do not have to be linked in the training reactions \cite{schwaller2019molecular, schwaller2020predicting, vaucher2020completion}. 

All these reaction completion methods could benefit from improving the underlying representation of reactions they are using. The following paragraphs will focus on the most promising improvements, and we will discuss how the three methods presented will benefit from it.
\begin{figure*}[htbp!]
    \centering
    \includegraphics[width=0.7\textwidth]{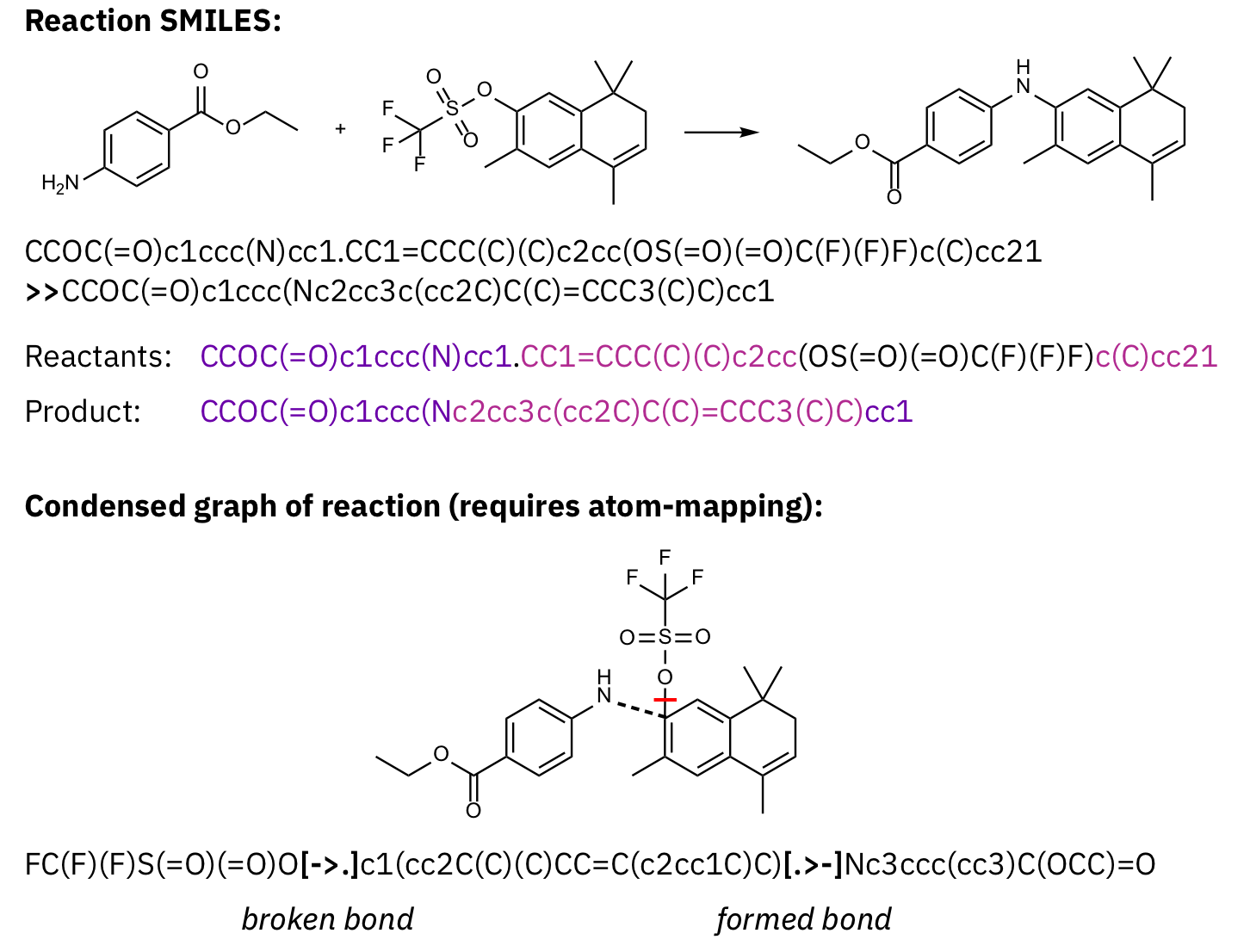}
    \caption{In most cases, the changes happening during the reaction affect only a small fraction of the molecule and everything else is left unchanged. However, current representations, like reaction \smiles, do not capture that, and major parts of the molecules are actually repeated. In contrast, condensed graphs of representation (CGRs) represent the bond changes in the reactions. To generate a CGR from a reaction \smiles, the atom-mapping has to be determined first. Agents and conditions are not shown in the figure.}
    \label{fig:reaction_figure}
\end{figure*}
The reactions present in the current data sets are rarely balanced, meaning not every atom from the left hand side of the chemical equation can theoretically be mapped to an atom to the right hand side. Indeed, in the literature, parts of a reaction are often omitted when they are either considered irrelevant or are unknown (for instance, not mentioning the side products), or so obvious it does not need to be mentioned (for instance, disregarding counterions or necessary byproducts such as CO\textsubscript{2}). While this makes sense when a human reads a reaction, since it improves clarity, it would be beneficial if the reactions were complete for an algorithm to learn from them. For graph-based methods, this would reduce the number of graph edits that need to be predicted as there would be less variation on both sides of the reaction. For text-based methods, this would allow a user to enforce an atom count at inference, which would most likely improve the performance. Finally, template-based methods would also benefit as the templates extracted from the data would be more consistent. 

A way to enforce the atom count of a reaction would be to describe only one side of the reaction, for instance the reactants, and then describe only the changes happening during the reaction \cite{hoonakker2011condensed}. This would not only enforce balanced reactions, but also remove the unnecessary redundancy of the current representation as illustrated in Figure~\ref{fig:reaction_figure}. Bort et al.\ \cite{bort_discovery_2021} proposed the use of such a text-based condensed graph of reaction (CGR) representation to perform property prediction. Extra symbols were added to the reactants to describe the reaction. This representation is well-suited for template-based methods as it turns every reaction into a ready-to-use template. This would also be convenient for graph-based methods as there is no need to extract the graph edits. Further work is required to make this kind of representation useful for text-based methods. The application of such methods is difficult if there is no separation between the changes and the initial molecules, which to some extent also applies to graph-based methods.

However, the atom-mapping that enables extracting reaction templates or graph edits and building CGRs is typically not directly available for experimentally observed reactions. Moreover, human labeling is prohibitively time-consuming for large databases. Traditionally, automated atom-mapping was performed using extended-connectivity-, maximum common substructure-, and optimization-based approaches \cite{chen2013automatic}. Schwaller et al.\ \cite{schwaller2021extraction} recently showed that accurate atom-mapping could be learned from reactions represented as \smiles without existing atom-mapping, through unsupervised training.

So far, we have discussed methods to improve the representation, but have not considered extending \selfies to represent reactions. We will consider two cases: a representation that is syntactically robust and one that is semantically robust. A syntactically robust representation would ensure the validity of the graph edits proposed. However, this would not guarantee that the results make sense chemically. This is the goal of the semantically correct representation. In the following project, we will discuss the benefits and the feasibility of such a representation.

\stepcounter{TaskCounter}
\SkipTocEntry\subsection*{Future Project \arabic{TaskCounter}: Graph edit rules and meta\selfies for reactions}
A syntactically robust reaction representation would most likely improve the performance of predictive models, as it is no longer possible to predict an invalid representation or an invalid graph edit sequence. To achieve this representation, the rule set that defines \selfies has to be extended significantly. Although they are significantly more comprehensive, it should still be possible to write down the set of rules corresponding to the possible graph edits.

A semantically robust reaction representation will be harder to achieve. The number of rules needed is probably extremely high. Our best estimate of the number of rules needed is from the work of Reference \cite{szymkuc_computer-assisted_2016}, with over 50,000 rules. Applying a similar approach to reaction \selfies will be quite an endeavor and will not be scalable, as the number of rules is too high. A more suitable approach would be to extract the rules from the data directly. Such rules could either be extracted using hand-crafted algorithms (similar to the project on meta\selfies for organic molecules), or could be learned with ML. The latter case requires the extraction of rules from the ML model, which could be achieved with symbolic regression of a trained neural network. This project is conceptually related with the project for molecules with \textit{complicated bonds}.

\section{Strings as programming languages}
\begin{figure*}[htbp!]
    \centering
    \includegraphics[width=1\textwidth]{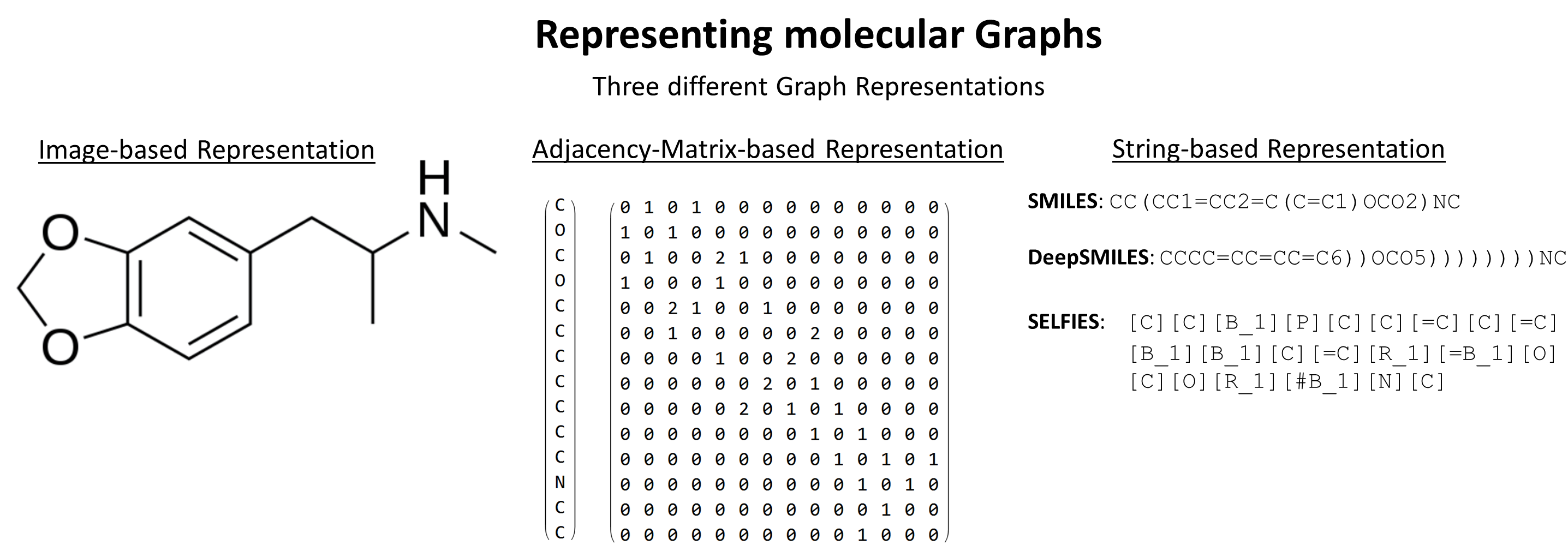}
    \caption{Graphs can be represented in numerous ways. For example, using images, adjacency matrices or strings. All of them are \textit{graph representations}. By relating string-based representations to programming languages, we show that they are in general the most expressive representations. For \selfies, \texttt{B1} and \texttt{R1} are appreviations for \texttt{Branch1} and \texttt{Ring1}, respectively.}
    \label{fig:three_rep}
\end{figure*}

String representations such as \smiles or \selfies are often considered less expressive and powerful than true ``graph-based'' representations, for instance those used in GNNs. However, fundamentally, quite the opposite is true for two very appealing reasons:

\begin{itemize}
\item \textbf{Strings and matrices can represent graphs:} Often, ``graph-based'' representations are understood implicitly as adjacency matrices. However, graphs are abstract objects, and can indeed be represented in diverse ways, for example by adjacency matrices, but also by strings (or other ways such as images). In that sense, both strings and matrices can be representations for graphs.
\item \textbf{Strings can store Turing-complete programming languages:} In the most general case, one can store the source code of computer programs as strings. For example, a Python file is a simple string, which is executed by the Python interpreter. Python is of course a Turing-complete language, which means, strings can encode the most powerful computational algorithms. Coming back to graph representations: one can imagine that \smiles or \selfies are programming languages, which are executed by an interpreter (for instance, by RDKit). The output of the program is a graph.
\end{itemize}

Arguably, \smiles and \selfies are rather simple programming languages, but this way of thinking indicates that one can develop much more powerful string-based molecular graph representations. These new molecular programming languages can be Turing-complete, thus encode arbitrary properties of a molecule that can be encoded in a computer. What follows now are a number of interesting future research questions that study the consequences of these ideas.

\stepcounter{TaskCounter}
\SkipTocEntry\subsection*{Future Project \arabic{TaskCounter}: Introducing ``molecular programming languages''}
Besides the performance of current string-based representations, the question remains how to extend string representations or \selfies to incorporate more prior information without losing desirable properties such as robustness. In the following, we propose two possible extensions to \selfies:

\begin{itemize}
    \item \textbf{Including 3D information such as bond angles and dihedral angles:} By incorporating 3D information, a \selfies could directly map to a specific molecular conformer, which could be beneficial in structure generation and embedding methods \cite{lemm2021machine}. In practice, extensive conformer searches could be circumvented if a specific configuration is already defined in a \selfies. A possible implementation of such 3D-\selfies could be envisioned through the use of pointer variables that locate positions in memory.
    \item \textbf{Including meta-characters for loops and logic:} Another important extension would include basic expressions of programming languages which can be used to enable different types of logic such as for loops to repeat substructures or characters for symmetric branches. Such characters could be of immense value to generate \selfies for larger and more complicated molecules (such as polymers or crystals, as discussed in previous chapters). The general idea of meta-characters goes hand-in-hand with the creation of a general purpose and domain-independent representation (i.e., meta\selfies), discussed in the Future Project 1.
\end{itemize}

\stepcounter{TaskCounter}
\SkipTocEntry\subsection*{Future Project \arabic{TaskCounter}: A 100\% robust programming language}
The discussion in the previous section motivates another leap: the possibility of a Turing-complete programming language that is 100\% robust, i.e.,\ every combination of elements in the instruction set gives a valid computer program. The question of deep generative models for code generation has just recently seen impressive progress in \mbox{OpenAI's} Codex, a GPT language model clone that was trained on all Python code on GitHub \cite{chen2021evaluating}. It would be exciting to explore possibilities for generative ML models that have access to a scripting language that produces valid code in every instance. Interestingly, the question of robust programming languages has been discussed in the field of artificial life since the pioneering 1993 work of Tierra \cite{ray1993evolutionary, adami1998introduction}. Extensions of these ideas have since been applied to studies on artificial evolution \cite{lenski2003evolutionary,wilke2001evolution}. We hope inspiration can be taken from that field of study.

\section{Comparing strings, adjacency matrices and images as molecular graph representations for ML}
Strings may be graph representations in the same way as adjacency matrix representations or image-based representations (cf. Figure~\ref{fig:three_rep}). Since strings are directly related to programming languages, they are in general the most expressive of all graph representations. A very important question is how these different graph representations differ in actual ML applications.

To answer this, it is interesting to note that different representations are suitable for different, specialized neural network architectures. Image-based representations can benefit from Convolutional Neural Networks (CNNs), adjacency matrix- based representations are the foundations for GNNs, and string-based representations work well for language models such as recurrent neural networks (RNNs) and Transformers.

The question of how these representations and their related ML models compete in the same task is so far underexplored. One very recent study has shown that chemical language models (using \selfies) and RNNs are powerful enough to generate very complex molecular distributions, including the largest molecules from PubChem \cite{flam2021keeping}. So far, GNN-based generative models struggle with this task and do not yet scale to these large sizes.

The comparison between the representations (and their corresponding models) leads to a number of interesting questions:
\begin{itemize}
\item \textbf{Memory footprint:}
As vehicles for storing molecular data, both strings and matrices should provide characteristic descriptions of the data. A fundamental principle for data description in ML is minimal description length (MDL). That is, the best description of the
data is given by the model which compresses it best. One example of MDL is Kolmogorov complexity \cite{kolmogorov1963tables}, which is defined as the length of the shortest computer program that produces the sequence of data. Even though Kolmogorov complexity itself is not computable, practical approximations of Kolmogorov complexity can be used to quantify the memory footprint of the molecular representation. This is especially important when using the strings or matrices as input to downstream algorithms for molecular property prediction or molecular generation. The level of physical memory burden incurred from using different representations can have significant impact on the execution speed, processor utilization, and energy cost of the program.
\item \textbf{Optimization difficulty:} Even if representations have the same memory footprint, their impact on the outcome of the ML algorithms may still vary. One reason is the difficulty of non-convex optimization. The resulting deep learning model may not be able to fully exploit the information in the data. The choice of input representation may also have an effect on the loss landscape of the neural network optimization problem, which would certainly influence training dynamics. Different molecular representations could lead to distinct local optima, producing models that differ in terms of generalization performance and sensitivity to input perturbation.
\item \textbf{Computational efficiency:}
From a computational perspective, string \textit{vs}.\ graph representation can also have different complexities due to the differences in numerical algorithms. For example, for strings of different lengths, one can either use sequential processing models such as RNNs, or Transformers with padding, which can be easily parallelized. However, the padded strings would have different sparsity structures (the patterns of zeros) than the matrix representations. These sparsity structures can be utilized to a varying degree in order to accelerate numerical operations including addition, multiplication or eigenvalue decomposition. The efficiency of the entire program, thus, can be easily affected.
\end{itemize}

To shed light onto these different properties, we suggest the following project.

\stepcounter{TaskCounter}
\SkipTocEntry\subsection*{Future Project \arabic{TaskCounter}: Comparisons in various data regimes in a regression task
}
While string-based representations tend to be more expressive and easier to generate, adjacency matrices in conjunction with GNNs have important advantages, such as permutation invariance. Images of the molecular graphs (which can be understood as another graph representation) could take advantage of extremely efficient, pretrained CNNs. A suitable experiment could be a discriminative task in the various data regimes. This of course depends on the target property to be learned. For example, for learning coordinate-dependent properties, it is still unknown how much prior information is actually necessary and whether string-based representations will outperform graph-based representations in the high data regime for specific tasks.

We suggest the development of a benchmark to compare image, adjacency matrix, and string representations for graphs in various data regimes for discriminative tasks. The PCQM4M-LSC data set may be useful for these comparisons: with approximately 3.8 million molecules and their associated HOMO-LUMO energy gaps (as estimated by DFT simulation), it poses a formidable chemical regression task \cite{nakata2017pubchemqc,wu2018moleculenet}.

The comparison should measure all three models in (at least) the prediction quality over the following characteristics:
\begin{itemize}
  \item the number of training epochs.
  \item the number of model parameters.
  \item various numbers of examples in the training data.
  \item various sizes (measured in edges) of the largest molecules in the training data set.
\end{itemize}

These experiments will give insightful answers about the characteristics of different data modalities in ML tasks, and will give experimental evidence about which models should be used in which situations in future practical applications.

\stepcounter{TaskCounter}
\SkipTocEntry\subsection*{Future Project \arabic{TaskCounter}: Comparisons in generative tasks}
A main motivation of \selfies is its application in generative, inverse-design tasks. We therefore suggest the development of new generative model benchmarks. For that, a number of important precautions need to be considered. First, when \selfies is used, a comparison among models based on their ability to generate valid molecules is no longer a useful design objective \cite{polykovskiy2020molecular}. Interestingly, previously used benchmarks \cite{polykovskiy2020molecular, brown2019guacamol} have also placed great importance on distributional learning metrics. However, this approach is reported to have multiple flaws in the form of edge cases \cite{renz2019failure}. For instance, simple algorithms that place carbon atoms at random positions within molecules have been shown to perform well on distribution matching objectives. Additionally, the recent proposal of the STONED algorithm \cite{nigam2021beyond}, which makes use of random \selfies mutations, has demonstrated ease in matching the structural distribution of molecules. FastFlows \cite{frey2022fastflows} uses normalizing flows to model distributions of molecules represented as \selfies and achieve fast sampling speeds. Another class of methods used for comparing molecular generative models can be classified as goal-directed benchmarks. In these, generative models compete among one another to optimize one or more molecular property functions.  It can also be important to generate dense \textit{local} chemical spaces, for example to create counterfactuals to explain black-box models \cite{wellawatte2022model}. Many of these tasks are provided within GuacaMol \cite{brown2019guacamol}: however, given the current rise of more sophisticated models, these benchmarks have become outdated. Recently, many generative models have been able to achieve perfect results on many of the GuacaMol tasks \cite{nigam2021janus, ahn2020guiding, winter2019efficient}, making it difficult to establish comparisons between models. Therefore, to compare deep generative models, one needs more sophisticated objectives that reflect the complexity of real-world molecular design. 

\section{Interpretability and usability of string-based representations}
\subsection{For humans}
Historically, representations have been developed with humans in mind for reading and writing molecules. String-based representations are more difficult to interpret than images of molecules, and an important question is their \textit{understandability} for humans. On the one hand, human chemists might want to write molecules quickly as text instead of drawing them, might be able to get a quick understanding of the structure without inserting it into a plotting tool, or might be interested in identifying substructures. On the other hand, readability for humans might not always be necessary. For example, \inchi strings are broadly used despite the fact that the human readability was considered to be of low importance when \inchi was designed \cite{heller2015inchi}. It is also worth pointing out that, while human readability is one of the often-cited advantages of \smiles, figuring out what a \smiles actually stands for can require significant intellectual effort. We just have to look at the \smiles for a simple steroid such as testosterone to see that this is the case:

\noindent\hspace*{0.5cm}\lstinline{O=C1CC[C@]2(C)[C@@]3([H])CC[C@]4(C)[C@@H](O)CC[C@@]4([H])[C@]3([H])CCC2=C1}

\noindent This suggests a trade-off in the necessity of readability and concrete computational applications. However, there is certainly a natural question of how well humans can \textit{interpret} molecular string representations, which has not been investigated experimentally to the best of our knowledge. Therefore, we suggest the following project.

\begin{figure*}[htbp!]
\centering
\includegraphics[width=0.95\textwidth]{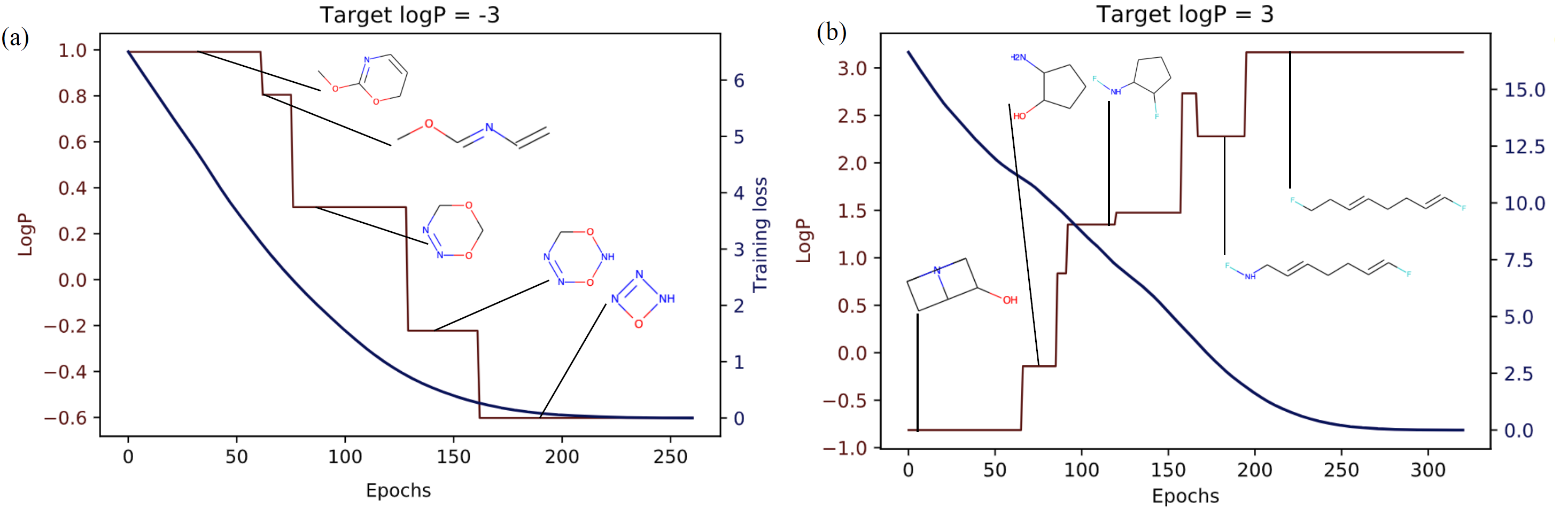}
\caption{\textbf{Pasithea, the DeepDreaming Generative Model} -- While the model continuously decreases the loss, the molecule changes in discrete steps. The target property was log\textit{P} of the molecule. The network is able to increase or decrease the molecular property almost steadily, which indicates a certain ``understanding'' of the representation. Image from \cite{shen2021deep}.}
\label{fig:deepdream}
\end{figure*}

\stepcounter{TaskCounter}
\SkipTocEntry\subsection*{Future Project \arabic{TaskCounter}: Experiment on readability of molecular string representations}
We suggest an experiment that tests the human readability of \mbox{\smiles-}, \mbox{\deepsmiles-}, \mbox{\selfies-}, and adjacency matrix-based representations of molecules. We envision a study with 50 or more participants from different countries. None of the participants must be previously familiar with these representations, to guarantee a fair comparison. The participants will get instructions for understanding each of the representations, with which they should familiarize themselves before the experiments start.

At the evaluation phase, the participants are asked to solve a number of tasks, such as substructure identification and translating the representation from and to molecular graphs. The participants will also be asked to solve some tasks in which they need to actively choose their preferred representation(s).

The results might help us to understand which representations are easiest to read, by analysing the accuracy, speed, and participant's preference of representations. Post-hoc interviews could then elaborate on the challenges of different representations and might help to design a potential \textit{Esperanto for Chemistry} -- an easy to understand language for molecules. Beyond human readability, such an experiment might allow us to compare and contrast which properties of representations are challenging for humans compared to computers. These results could potentially lead to interesting findings on the differences between humans and machines, thus showing where we should place our trust in our intuitions around ML for chemistry.

\subsection{For machines}
An interesting question is how ML models interpret different representations. Specifically, if \selfies is used in a generative model, all generated molecules are correct. In this case, how can one be sure that the model's output is meaningful concerning some metrics such as usefulness and not just a collection of random strings which, by construction, lead to valid molecules? Furthermore, how can the machine interpretability of different representations be compared, specifically between \smiles and \selfies? In other words, which one is ``easier'' to learn for machines?

In deep generative models using VAEs, the latent space using \smiles consists of numerous, scattered, valid regions that exist within invalid valleys (see Figure~\ref{fig:selfies_latentspace}). In contrast, the entire latent space corresponds to valid molecular structures if \selfies is employed instead. This fact allows for the application of continuous gradient descent optimization in the latent space, where the optimizer will always provide meaningful structures. The robustness, however, does not necessarily correspond to a smooth encoding in the latent space, per se, where small changes in the latent space lead to small modifications in the molecule. Therefore, it remains to be seen whether generative models can actually learn structure-property relations using \selfies.

\textbf{Deep Molecular Dreaming} -- One experiment that tackles the problem of interpretability and smoothness to a certain extent employs the technique of DeepDreaming \cite{shen2021deep}. The generative model denoted as Pasithea consists of a single neural network that is used for the generation of molecules in two steps. In the first of these, the network learns to predict a chemical property given a one-hot encoding of a \selfies. In the second step, the neural network weights are frozen and a target value of the property is fixed. Gradient descent is then used with respect to the one-hot encoding, meaning that the input molecule is continuously modified. The results of two design-processes are shown in Figure~\ref{fig:deepdream}. While the model continuously decreases the loss, the one-hot encoding of the molecule is changed within the discrete space. It is apparent that the target property increases/decreases for positive/negative target values of log\textit{P} in a nearly monotonous way. This indicates that the model has indeed understood an essence of the structure-property relation, and is not exploiting only the robustness of \selfies. A complementary approach is to use directly invertible neural networks for generative models, such as presented in \cite{hu2021inverse}.

\begin{figure*}[htbp!]
\centering
\includegraphics[width=0.7\textwidth]{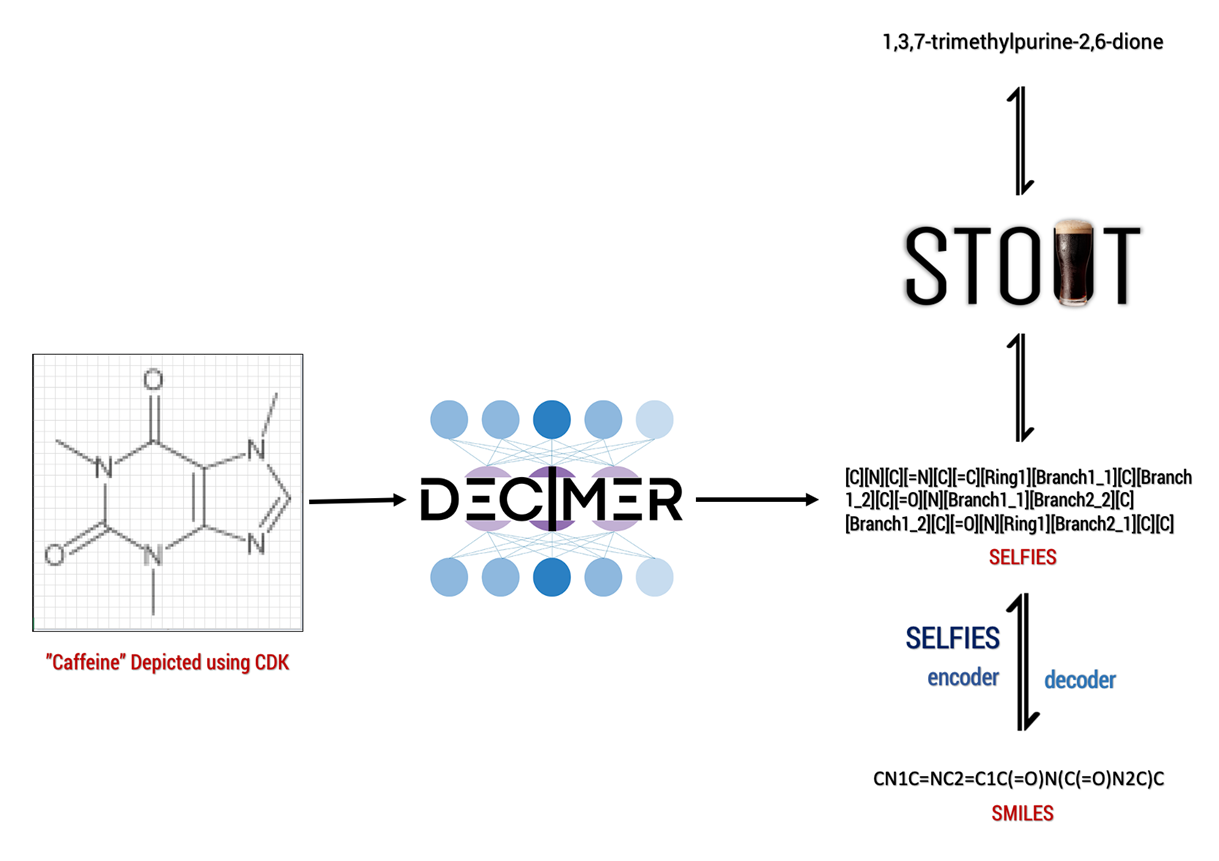}
\caption{\textbf{DECIMER \& STOUT} -- A framework for translating images or strings to \smiles. Experiments show that the application of \selfies as an intermediate representation improves the results, which indicates that ML models find it easier to read and write \selfies compared to \smiles. These indications are surprising because it is not clear how the model exploits \selfies's robustness to improve results. Image from \cite{rajan2020decimer,rajan2021stout}.}
\label{fig:decimer}
\end{figure*}

\textbf{DECIMER} -- Optical Chemical Structure Recognition (OCSR) tools have been developed to extract chemical structures and convert them into a computer-readable format. The best-performing OCSR tools are mostly rule-based algorithms. To address the OCSR problem by using the latest computational intelligence techniques and provide an automated open-source software solution, DECIMER (Deep lEarning for Chemical ImagE Recognition) was launched Figure~\ref{fig:decimer} \cite{rajan2020decimer}. One of the biggest challenges in developing DECIMER was to use the string representation of chemical structures in a meaningful way. The issues encountered initially with \smiles were splitting them into meaningful tokens during training and evaluation, when the predicted \smiles were syntactically and semantically incorrect, reducing the accuracy of the tool. As a result of using \selfies, these issues were resolved, leading to better training of models. Additionally, it demonstrates how efficiently neural networks can be trained to read and write \selfies strings.

\textbf{STOUT} -- A conceptually related tool is STOUT (\smiles-TO-IUPAC-name Translator). It was developed to translate between the IUPAC names and string-representations of molecules. IUPAC developed a naming scheme for chemistry based on a set of rules. Due to the complexity of this rule set, assigning a chemical name is challenging for humans, and there are a limited number of rule-based cheminformatics applications available to assist with this process, all of which are commercial. STOUT is an open-source, deep-learning-based neural machine translation approach developed to generate the IUPAC name for a given molecule from its \smiles string and carry out the reverse translation \cite{rajan2021stout}. One key observation was that STOUT works better when using \selfies as an internal representation than with \smiles. Therefore, the \smiles strings are internally converted into \selfies before the input is processed by the model. Likewise, the predicted \selfies are decoded back into \smiles during reverse translation. This is another indication that \selfies is \textit{understood better} than \smiles for some complex deep learning tasks.

\textbf{\selfies in a language model} -- It was shown recently that an RNN language model trained on \selfies is more robust to over-fitting than with \smiles \cite{flam2021keeping}. This is understood from the larger novelty of the generated molecules at similar quality of the learned distribution. 

There are numerous future experiments which could shed light into the ``understandability'' of different representations. We summarize a few of them:

\stepcounter{TaskCounter}
\SkipTocEntry\subsection*{Future Project \arabic{TaskCounter}: Translation between different types of representations}
It would be interesting to train a neural network which can translate between different representations of molecular graphs, including (current or future) string-based representations, adjacency matrix representations or images of molecular graphs. This would be exciting for two reasons. Firstly, if the neural network learns to work with three entirely different representations, it might build up an interesting and robust internal representation which could be analysed subsequently. Secondly, it gives the opportunity to combine three of the most powerful ML methods at the same time, namely GNNs for the adjacency matrix representation, Transformers for strings, and CNNs for the images of molecular graphs. A concrete use case could look like this: the goal is to predict a molecular property from a molecule that is encoded as a \selfies. The neural network translates the \selfies to an adjacency matrix and an image, producing a latent meta-representation of the molecule in one of its hidden layers in the process. All or some of these four representations are provided to downstream models with appropriate architectures (e.g., GNN for an adjacency matrix or Transformer for a string), which are then ensembled to produce better predictions and overcome deficiencies in each individual chemical representation. Note that some important progress has already been achieved in translation tasks. Examples are image to string representation translations \cite{rajan2020decimer,clevert2021img2mol}, and string to IUPAC \mbox{translations \cite{winter2019learning,rajan2021stout}}.

\stepcounter{TaskCounter}
\SkipTocEntry\subsection*{Future Project \arabic{TaskCounter}: Which string-based representations allow for simpler models and faster training?}
Several experiments could be performed to determine how the use of different representations for training ML models on the same set of regression tasks impacts learning and final quality metrics, such as accuracy. Initially, these tasks should comprise the usual benchmark endpoints for ML prediction, such as boiling points, log\textit{P} and \emph{pK}\textsubscript{a}. In addition, tasks known to be influenced by the 3D structure of the compounds, such as predicting HOMO or LUMO energies, or activity toward a biological target, could also be explored. 

In a first experiment, models with the same end goal could be trained to determine how different representations impact the final accuracy and how they impact the model's ability to achieve better performance with less training time. In another experiment, the numbers of neurons and layers of neural networks would be decreased, and the number of episodes necessary to reach a certain quality would be recorded. This project would allow us to verify the ability of models trained on \selfies to generalize better, provided the performance after these model simplifications does not decrease as fast as for models trained on different representations.

\stepcounter{TaskCounter}
\SkipTocEntry\subsection*{Future Project \arabic{TaskCounter}: Smoothness of generated molecules}
Another interesting experiment would be to investigate the smoothness of latent spaces of VAEs trained with \smiles, \deepsmiles, and \selfies. If one wants to use gradient-based optimizers in the latent space, it would be desirable if the properties of the generated molecules changed to a small extent when sampling from closely related points in the latent space. We suggest to measure a set of properties for each generated molecule while continuously wandering in the latent space. Notably, the design of such an ML experiment needs to take the invalid regions of the latent space into account.

\stepcounter{TaskCounter}
\SkipTocEntry\subsection*{Future Project \arabic{TaskCounter}: Learning what the machine has learned in the latent space}
The latent space represents the intrinsic representation that has been learned by the model to solve a given task. It will be exciting to understand what this representation stands for. If one understands how a VAE encodes and decodes molecules to and from the latent space, some of the questions presented above can likely be answered even without performing further experiments. To that end, t-SNE \cite{van2008visualizing} and other dimensionality reduction tools are expected to be challenging to interpret, thus one direction could be the applications of latent spaces with only two or three dimensions, which can be displayed without projections. Related projects have rediscovered interesting physical concepts such as the heliocentric coordinates \cite{iten2020discovering}, the arrow of time, \cite{seif2021machine} or interpretation in quantum optics \cite{krenn2020computer,flam2021learning}, and we expect similar exciting possibilities in material science and chemistry.

\section{Conclusion}
The resolution of the 16 proposed challenges could significantly advance the applicability of AI in diverse fields of chemistry and beyond. Furthermore, questions about the interpretability of languages for machines could help us understand how a machine solves complex tasks in chemistry -- what principles or concepts it uses. This could be a path for human scientists to learn ideas from AI in chemistry. We hope that our journey of possibilities will inspire researchers in the cheminformatics and applied AI community and lead to exciting new results and advances in molecular string representations.

\SkipTocEntry\section*{Acknowledgements}
The authors thank Greg Landrum and Daniel Flam-Shepherd for valuable comments on the manuscript.

The authors also thank Sara Bebbington of IOP Publishing and Zamyla Chan and Erin Warner of the University of Toronto Acceleration Consortium for helping to organize the \selfies workshop.

M.K.\ acknowledges support from the FWF (Austrian Science Fund) via the Erwin Schr\"odinger fellowship No.\ J4309.

R.F.L.\ receives a PhD Scholarship from the S\~{a}o Paulo Research Foundation (FAPESP) -- Grant \#2021/01633-3. This study was financed in part by CAPES -- Finance Code 001.

R.P.\ acknowledges funding through a Postdoc.Mobility fellowship by the Swiss National Science Foundation (SNSF, Project No.\ 191127).

A.W.\ would like to thank the Natural Sciences and Engineering Council of Canada (NSERC) for financial support via a CGS-M scholarship.

G.T.\ acknowledges financial support from NSERC via the PGS-D scholarship.

R.Y.\ acknowledges support from U.S.\ Department Of Energy, Office of Science, AWS Machine Learning Research Award, and NSF Grant \#2037745.

D.L.\ and G.v.R.\ were supported by the von Lilienfeld lab at the University of Vienna.

A.D.W.\ was supported by the National Institute of General Medical Sciences of the National Institutes of Health under award number R35GM137966.

K.M.J.\ and B.S.\ acknowledge funding from the European Research Council (ERC) under the European Union’s Horizon 2020 research and innovation programme (grant agreement No.\ 666983, MaGic).

J.M.N.D.\ acknowledges support by the National Council on Science and Technology (CONACYT) under award Number CVU 105568.

P.S.\ acknowledges support from the NCCR Catalysis (grant number 180544), a National Centre of Competence in Research funded by the Swiss National Science Foundation.

S.M.M.\ was supported by the Swiss National Science Foundation (SNSF) under Grant P2ELP2\_195155.

U.S.\ acknowledges support from the Deutsche Forschungsgemeinschaft (DFG) within NFDI4Chem (Grant No.\ NFDI4-1).

Q.A.\ acknowledges support from the National Science Foundation (Grant No.\ DMR-1928882).

A.A.-G. \ acknowledges support from the Canada 150 Research Chairs Program, the Google Focused Award, and Dr. Anders, G. Fr\o seth.

\bibliography{refs}

\end{document}